\documentclass[aps,prb]{revtex4}

\usepackage{graphicx,amssymb,epsfig}
\newcommand{\be}{\begin{equation}}
\newcommand{\ee}{\end{equation}}
\newcommand{\bea}{\begin{eqnarray}}
\newcommand{\eea}{\end{eqnarray}}

\newcommand{\bp}{\ensuremath{\mathbf p}}

\newcommand{\bq}{\ensuremath{\mathbf q}}
\newcommand{\br}{\ensuremath{\mathbf r}}

\newcommand{\Tr}{\mathop{\rm Tr}}
\newcommand{\re}{\mathop{\rm Re}}
\newcommand{\im}{\mathop{\rm Im}}

\usepackage{amsmath}
\usepackage{graphicx}

\newcommand{\e}{\varepsilon}

\newcommand{\sgn}{\mathop{\rm sgn}}
\begin{document}
\title{Interaction effects on magnetooscillations in a two-dimensional electron gas}
\author{ Y.~Adamov$^{1,2}$, I.~V.~Gornyi$^{2\dagger}$, and A.~D.~Mirlin$^{2,3\#}$}
\affiliation{
$^1$ Department of Physics, Brookhaven National Laboratory, Upton, NY 11973-5000, USA\\
$^2$Institut
f\"ur Nanotechnologie, Forschungszentrum Karlsruhe, 76021 Karlsruhe,
Germany
\\
$^3$Institut f\"ur Theorie der Kondensierten Materie, Universit\"at
Karlsruhe, 76128 Karlsruhe, Germany
}

\begin{abstract}
Motivated by recent experiments, we study the interaction corrections
to the damping of magnetooscillations in a two-dimensional electron
gas (2DEG).  We identify leading contributions to the interaction-induced
damping which are induced by corrections to the effective mass and
quantum scattering time.  The damping factor is calculated for Coulomb
and short-range interaction in the whole range of temperatures,
from the ballistic to the diffusive regime.  It is shown that
the dominant
effect is that of the renormalization of the effective electron mass
due to the interplay of the interaction and impurity scattering. The
results are relevant to the analysis of experiments on
magnetooscillations (in particular, for extracting the value of the
effective mass) and are expected to be useful for understanding the
physics of a high-mobility  2DEG near the apparent metal-insulator
transition.

\end{abstract}

\date{\today}

\maketitle

\section{Introduction}
\label{s1}

The influence of electron-electron interaction on  transport
properties of low-dimensional disordered conductors at sufficiently
low temperatures $T$ remains one of central topics of the condensed matter
physics. In a seminal set of works (see the review \cite{altshuler}),
Altshuler and Aronov studied the
effects of the interplay of interaction and disorder on conductivity
and tunneling density of states in the diffusive regime characterized
by the condition $2\pi T\tau\ll 1$, where $\tau$ is the transport mean
free time (we set $k_B=\hbar=1$). Their results were generalized
within the framework of the renormalization group (RG) by Finkelstein
\cite{finkelstein}.    The last decade has witnessed
a renewed increase of activity in this field, largely motivated by
experiments on an apparent metal-insulator transition in 2D systems.
This interest was triggered by experiments \cite{kravchenko94-96}
which showed a ``metallic'' behavior
(decrease of resistivity with lowering $T$) in high-mobility Si
structures. Later, qualitatively similar behavior was observed in a
variety of high-mobility 2D systems, see
Refs.~\onlinecite{abrahams01,altshuler01,kravchenko04,pudalov04a}
for  reviews.

The metallic behavior has been attributed to the effects of the
electron-electron interaction in the ballistic temperature range,
$2\pi T\tau \gg 1$. These effects were originally considered in the
framework of the temperature-dependent screening
\cite{temp-dep-screen}. More recently, a systematic theory was
developed, taking into account also exchange contributions and the
effects of both parallel and transverse magnetic fields, and valid in
the whole range of $T$ from the diffusive to the ballistic regime
\cite{ZNA,GM}. Another mechanism that can explain the metallic
behavior of resistivity in an intermediate temperature range in the
diffusive regime was studied within the RG framework in
Ref.~\onlinecite{punnoose}. It is applicable to systems with more
than one valley, such as silicon MOSFETs.

Despite these successes of the theory, numerous experimental observations
remain puzzling and wait for an explanation. In particular, it was
found that the spin susceptibility, proportional to the product
$mg$ of the effective mass $m$ and the
$g$-factor, is strongly growing when the density approaches the value $n_c$
corresponding to the apparent transition. This conclusion was drawn on
the basis of several experimental methods, including
the analysis of beating pattern of Shubnikov-de Haas
oscillations \cite{okamoto,kravchenko00,pudalov02,zhu}, the study of
magnetoresistance\cite{vitkalov01,shashkin01} in the parallel field,
and  measurement of thermodynamic magnetization \cite{prus03};
see recent reviews \cite{pudalov04b,shashkin05}.

The enhancement of susceptibility with lowering density, interpreted
in a number of papers as its divergence at $n=n_c$,
has attracted a great deal
of attention, since it might be an indicator of some phase transition
that the system undergoes with a decrease of density. The
interpretation of the data has remained, however,
controversial. In particular, it remained unclear whether the strong
increase of spin susceptibility should be attributed to that of $m$
or of $g$. This information is of crucial importance for understanding
the nature of the possible transition.

Several experimental approaches have been used to separate the
behavior of the effective mass from that of the $g$-factor.
In Refs.~\onlinecite{shashkin02,proskuryakov02,pudalov03}
a fit of the resistivity data
to the theoretical formulas of Ref.~\onlinecite{ZNA} was used to find the
interaction constant $F_0^\sigma$, and thus the $g$-factor. The
accuracy of this procedure is questionable, since the theory of
Ref.~\onlinecite{ZNA} neglects higher Fermi-liquid interaction constants and
assumes isotropic impurity scattering. Another approach is based on
thermodynamic measurements in strong magnetic field
\cite{anissimova05}. However, the authors of this work were able
to measure the effective mass in a very narrow interval of electron
concentration only, so that the results are not too informative. Also,
a strong magnetic field is expected to
influence strongly the characteristics of the electron liquid, so that
the applicability of such measurements to the low-field properties is
questionable. So, while most of the above measurements seem to indicate
that it is the effective mass that is responsible for the strongly enhanced
susceptibility, an independent verification is clearly needed.

A well known method for determination of the effective mass is based
on the investigation of the temperature dependence of Shubnikov-de
Haas oscillations (SdHO). It was applied to the present problem in
Refs.~\onlinecite{pudalov02,shashkin03}. However, the analysis of the SdHO data
is complicated  by the fact that both the effective mass and
the elastic quantum scattering time $\tau_q$
are $T$-dependent, in view of the combined
effect of interaction and disorder. An unambiguous interpretation of
experimental data requires \cite{pudalov02} a theoretical information
on $T$-dependence of $m$ and $\tau_q$. A development of the
corresponding theory is the aim of the present paper.

In fact, a recent paper \cite{MMR} has made an important step in this
direction. Specifically, it was shown in Ref.~\onlinecite{MMR} that the
Lifshitz-Kosevich formalism \cite{lifshitz},
originally developed for the analysis of
magnetooscillations in a 3D Fermi liquid, is also applicable in 2D in
the regime where the oscillations are exponentially suppressed by
temperature smearing or disorder. (In the regime of strong
oscillations, the Lifshitz-Kosevich formula in 2D should be modified, as
was earlier shown in Ref.~\onlinecite{stamp}.) Another result of
Ref.~\onlinecite{MMR} is
that the inelastic electron-electron relaxation does not contribute to
the damping of magnetooscillations (similarly to the earlier result of
Ref.~\onlinecite{fowler} for the case of electron-phonon scattering).

The authors of Ref.~\onlinecite{MMR} then 
calculated the contribution to
the damping induced by the interplay of interaction and disorder.
Their theoretical treatment of the problem is, however,
far from complete.
First, they consider only diagrams for the self-energy with one
impurity-ladder vertex correction to the interaction line and discard
diagrams with no and with two vertex
corrections.
Second, they claim that the $T$-dependence of the
oscillation damping rate can be
equivalently attributed either to the correction to the
effective mass, or to the
quantum scattering rate (Dingle temperature).
Furthermore, in the latter case
their result for the $T$-dependence of
$\tau_q$ is in contradiction with the picture of Friedel oscillations
inducing a correction to the relaxation rate, which is linear in $T$
and is governed by backscattering \cite{ZNA}.

In addition to the above experimental motivation, the development of
the theory of interaction effects on magnetooscillations in a
disordered 2DEG represents a fundamental theoretical problem.
Such a theory should complement the recently developed theory of
interaction effects in transport of 2D electrons in zero and
non-quantizing magnetic fields~\cite{ZNA,GM}.
Let us emphasize a peculiar aspect of the present problem.
The damping of oscillations is governed by the
self-energy, which is a single-particle quantity.
[Indeed, the relevant diagrams, see Sec~\ref{s2.2}  below,
are reminiscent of those for the tunneling density
of states (DOS)]. Generally, the self-energy is not a
gauge-invariant object.
On the other hand, the magnetization and the conductivity
(magnetooscillations of which we would like to study) are
observable (and thus gauge-invariant) quantities. It is well known that the
gauge-invariance is of crucial importance for interaction-induced
corrections; a difference between the results for conductivity and for
tunneling DOS in the case of Coulomb interaction
\cite{altshuler,finkelstein} serves as a nice illustration. It is thus
a theoretical challenge to see how the gauge invariance manifests
itself in the magnetooscillation problem.

The outline of the article is as follows. The section \ref{s2} is
devoted to presentation of the general formalism. In Sec.~\ref{s3} we
apply it to calculate the interaction-induced contribution to the damping of
magnetooscillations in the case of short-range interaction. In Sec.~\ref{s4} we
show how to extract from the above result the corrections to the
effective mass and the quantum scattering time. We also perform a
calculation of the correction to the scattering time based on the
picture of Friedel oscillations and demonstrate a complete agreement
between the two approaches. In Sec.~\ref{s5} we generalize our results
to the case of Coulomb interaction. We also perform there a comparison
with the calculation of Ref.~\onlinecite{MMR}. Section~\ref{s6}
summarizes our findings. Some technical details of our calculations
are presented in Appendices.

\section {Magnetooscillations: general formalism}
\label{s2}

\subsection{Derivation of the formula for a decay of the oscillations}
\label{s2.1}

We begin by calculating the oscillatory part $\Omega_{\rm osc}$ of the
thermodynamic potential $\Omega$. From this quantity one can derive
the oscillating contribution to the thermodynamic density of states
\begin{equation}
\frac{\partial n_{\rm osc}}{\partial \mu}=
\frac{\partial^{2}\Omega_{\rm osc}}{\partial \mu^2},
\label{compress}
\end{equation}
where $\mu$ is the chemical potential,
and de Haas-van
Alphen oscillations of magnetic susceptibility
\begin{equation}
\label{mag-susc}
\chi_{\rm osc}=-\frac{\partial^{2}\Omega_{\rm osc}}{\partial B^2},
\end{equation}
where $B$ is a magnetic field. The main subject of our interest is the
exponential damping factor of these magnetooscillations.
For non-interacting electrons, the same exponential damping factor governs the magnitude 
of the Shubnikov -- de-Hass oscillations of the conductivity for the case of weak
disorder potential in sufficiently weak magnetic field~\cite{ando,MPW}, where the self-consistent
Born approximation~\cite{ando} (SCBA) is valid.
As we are going to show, the interaction-induced correction to the damping factor of  
the thermodynamic density of states arises due to the renormalization of the effective mass 
and the quantum scattering time.
Therefore these $T$-dependent corrections to the damping factor govern 
the magnitude of the Shubnikov -- de-Hass oscillations as well, similarly to the
non-interacting case.

Our starting point is the
expression for the thermodynamic potential derived in the paper by Luttinger and Ward
\cite{LuttiWard60}
\begin{equation}
  \label{eq:Luttinger}
  \Omega=-T\Tr\ln(-G^{-1})-T\Tr (G \Sigma)+\Omega'.
\end{equation}
Here
\be
G(i\e_n,m\omega_c)=[G^{-1}_0(i\e_n,m\omega_c)-\Sigma(i\e_n,m\omega_c)]^{-1},
\ee
is the dressed Green's function in the Matsubara formalism,
$i\varepsilon_n=(2n+1)i\pi T$ is the Matsubara fermionic energy,
$\omega_c$ is the cyclotron frequency, and $m$ is the Landau level
index. Further,
$$
G_0(i\e_n,m\omega_c)=
\frac{1}{i\varepsilon_{n}+\mu-(m+1/2)\omega_c}
$$
is the Green's function in the absence of disorder and interaction,
and $\Sigma(i\e_n,m\omega_c)$ is a self-energy part of Green's
function which includes all the disorder and interaction effects.

\begin{figure}[h]
 \includegraphics[width=0.9\columnwidth]{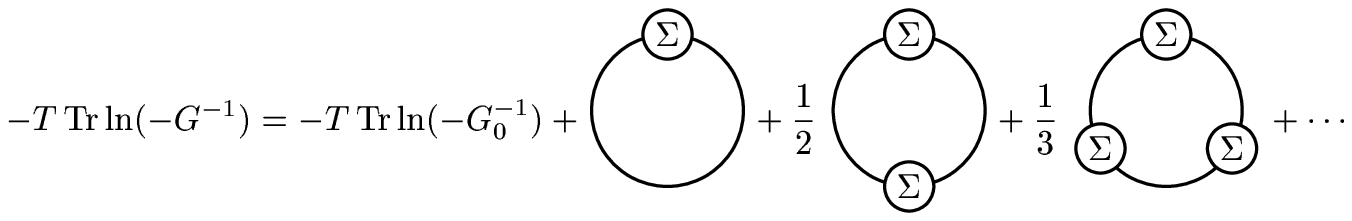}
 \caption{The logarithmic term $\tilde\Omega$, Eq.~(\ref{Omega-trace}),
 in the thermodynamic potential, Eq.~(\ref{eq:Luttinger}),
 is a sum of closed loop diagrams with self-energy insertions $\Sigma$.
 This term is responsible for magnetooscillations.\label{f-closed}}
\end{figure}

The trace in Eq.~(\ref{eq:Luttinger}) implies summation over Matsubara
frequencies $\e_n$ and over Landau levels $m$. The logarithmic
term contains all the closed loop diagrams with insertion of
self-energy~(Fig.~\ref{f-closed}). The terms $-T\Tr (G \Sigma)$ and $\Omega'$
are introduced to avoid double-counting of diagrams~\cite{LuttiWard60,AGD}.
The term $\Omega'$ denotes the sum of all so-called skeleton diagrams
with all bare Green's functions replaced by dressed Green's functions
(for the recent discussion of Luttinger-Ward formalism in 2D Fermi systems see
Refs.~\onlinecite{chubukov1,chubukov}).

As shown in Ref.~\onlinecite{luttinger}, the exponential decay of
magnetooscillations is described by the $\Tr\ln$-term. The oscillatory parts
of the additional terms, which are introduced  to fight
overcounting, cancel each other.
In order to obtain the correction to the thermodynamic potential
we need to calculate the self-energy part of the Green's function.

We decompose the self-energy into two parts:
\be
\Sigma(i\e_n,m\omega_c)=\Sigma_{\rm dis}(i\e_n,m\omega_c)+
\Sigma_{\rm ee}(i\e_n,m\omega_c).
\ee
where
$\Sigma_{\rm dis}(i\e_n,m\omega_c)
$
denotes the self-energy part due to the scattering on disorder potential with
electron-electron interaction switched off
and
$\Sigma_{\rm ee}(i\e_n,m\omega_c)
$
contains all the interaction effects.

In this paper we assume that disorder potential is $\delta$-correlated,
inducing a large-angle scattering of electrons.
The disorder-induced (noninteracting) part of the self-energy
for white-noise disorder and weak magnetic field
$\omega_c\tau\ll 1$ is given by
\begin{equation}
\Sigma_{\rm dis}(i\e_n,m\omega_c)={i\sgn\e_n\over 2\tau}.
\end{equation}
For stronger magnetic field (i.e. for separated Landau levels),
one should employ the SCBA. In this case, both the real and
imaginary parts of the self-energy depend on $i\e_n$ in a non-trivial way.
In this paper, however, we will address only weak magnetic
fields when Landau levels overlap.

We thus consider the relevant term ${\tilde\Omega}=-T\Tr\ln(-G^{-1})$
in the thermodynamic potential,
\be
  {\tilde\Omega} =-2\nu\ T\sum_{n=-\infty }^{\infty }\omega_{c} \sum_{m=0}^{\infty} \
   \ln \left[
    \omega_{c}m-\mu -i\varepsilon _{n}+
    \Sigma(i\e_n,\omega _{c}m)
  \right].
  \label{Omega-trace}
\ee
For overlapping Landau levels,
\begin{equation}
\label{overlap}
\omega_c\tau\ll 1,
\end{equation}
the $k$-th harmonics of the magnetooscillations 
$$
\Omega_{\rm osc}=\sum_k A_k \cos\frac{2 k \pi^2 n_e }{eB}
$$
is damped by disorder even at zero
temperature via
the standard Dingle factor $\exp(-\pi/\omega_c\tau)$.
Therefore we will consider only the first harmonics of the oscillations, $A_1$,
neglecting all $A_k$ with $k>1$ (whose damping is much stronger).

The oscillatory part of (\ref{Omega-trace}) is calculated in Appendix A:
\bea
\Omega_{\rm osc}
&\simeq&
2\nu \left({\omega_c\over 2\pi} \right)^2
A_1\cos{2\pi^2  n_e \over e B},
\label{eq-omega-osc}
\eea
($n_e$ is the electron concentration) with the amplitude of the first harmonics of the oscillatory part of the
thermodynamic potential
given by
\bea
A_1&\equiv&\frac{4\pi^2 T}{\omega_c}\sum_{\e_n>0}
  \exp\left(
    -\frac{2\pi}{\omega_c^*}
    \left[
    \e_n+{1\over 2\tau(1+\alpha_0)} +
      {i\delta\Sigma (i\e_n,\xi_0)\over 1+\alpha_0} \right]
  \right).
  \label{A1-trace}
\end{eqnarray}
Here $\delta\Sigma (i\e_n,\xi_0)$ is the self-energy part
related to the interplay of disorder and interaction. It is analytically continued
from the points $m\omega_c$ to the whole complex plane $\xi$ and taken
at $\xi=\xi_0$, where $\xi_0$ [defined in Eqs.~(\ref{xi00}) and (\ref{xi0result})
of Appendix A] is the pole of the  Green's function in
the presence of disorder.

The coefficients $\beta_0$ and $\alpha_0$ determine the Fermi-liquid
(FL) renormalization of the effective mass in a pure system at zero $T$,
\begin{equation}
m^{*}=m\frac{1+\alpha_0}{1+\beta_0}.
\label{m*}
\end{equation}
The effective mass $m^*$ in turn governs  the expression for
the FL-renormalized effective cyclotron frequency,
\begin{equation}
\omega_c^*={e B\over m^*}=\omega_c\frac{1+\beta_0}{1+\alpha_0}.
\end{equation}
The coefficient $\alpha_0$ is related to the FL renormalization of the
$Z$-factor,
\be
Z=\frac{1}{1+\alpha_0},
\label{Zfactor}
\ee
which is given by the residue of the Green's function.

Depending on the relation between temperature $T$ and the
elastic scattering rate $1/\tau,$ there are two regimes:
ballistic, $T\tau \gg 1$, and diffusive, $T\tau \ll 1$
(more accurately, the relevant dimensionless parameter is $2\pi T\tau$).
In the ballistic regime, it follows from Eq.~(\ref{overlap})
that $T\gg 1/\tau \gg\omega_c$. The diffusive regime can be further split into
two subregimes: normal diffusive ($\omega_c\ll T\ll 1/\tau$) and
ultra-diffusive ($T\ll\omega_c\ll 1/\tau$).
When $T\gg\omega_c$, as in the ballistic and normal diffusive regimes,
only the first Matsubara frequency $\e_0=\pi T$ in the sum determining $A_1$ is relevant,
since the contribution of higher Matsubara frequencies are exponentially suppressed.
On the other hand, in the ultra-diffusive regime $T\ll\omega_c$ and higher Matsubara
frequencies contribute as well.

In what follows we concentrate on the case $T\gg\omega_c$.
Under this condition, we get
\be
A_1={4\pi^2 T\over \omega_c}\exp\left[-{2\pi\over \omega_c^*}\,
\left\{\pi T + i\, Z\, \delta\Sigma(i\pi T,\xi_0)\right\}\right]
\ \exp\left[-{\pi\over\omega_c^*\tau^*}\right],
\label{A1}
\ee
where we introduced
the FL-renormalized scattering time
\begin{equation}
\tau^*=\tau\, (1+\alpha_0).
\label{taustar}
\end{equation}
We note that this renormalization of $\tau$ is incomplete
since
it does not include the FL vertex corrections to the impurity scattering
line. The corresponding contributions is contained in $\delta\Sigma(i\pi T,\xi_0)$
[$T$-independent terms in
Eqs.~(\ref{delta-sigma-vert-short-result}) and (\ref{coulomb-delta-sigma-result})] and
will be addressed in Sections III and IV
below.
We also note that the inelastic contribution to the self-energy
$\propto [(\pi T)^2-\e_n^2]\sgn\e_n$ (see Appendix B)
vanishes for $\e_n=\pi T,$ and thus does not affect the damping of the
magnetooscillations $B(T)$ for $T\gg\omega_c,$
in agreement with Refs.~\onlinecite{MMR,fowler}.

Using the renormalized quantities in Eq.~(\ref{A1}) we represent $A_1$
in the form
\begin{equation}
A_1=A_1^{(0)}(T)\, \exp[B(T)],
\end{equation}
where
\begin{equation}
A_1^{(0)}(T)\equiv {4\pi^2 T\over \omega_c}\exp\left[-{2\pi^2 T\over\omega_c^*}
- {\pi\over\omega_c^*\tau^*}\right]
\label{A10}
\end{equation}
is the standard FL Lifshitz-Kosevich result
and
\begin{equation}
\label{FT}
B(T)= -{2\pi i \, Z \, \delta\Sigma(i\pi T,\xi_0)\over \omega_c^*}.
\end{equation}
We thus see that in order to obtain the additional interaction-induced
damping factor of magnetooscillations it is necessary to evaluate
$Z\, \delta\Sigma(i\pi T,\xi_0)$.

\subsection{Self-energy}
\label{s2.2}

We begin by considering the interaction-induced self-energy part
$\Sigma_{\rm ee}(i\e_n,m\omega_c)$ in the lowest order in interaction.
This is sufficient in the case of a weak short-range interaction
analyzed in Sec.~\ref{s3} below. For the more realistic case of the
Coulomb interaction (Sec.~\ref{s5}),
the relevant higher-order terms can be treated
using the random-phase approximation (RPA).
Higher-order contributions to the $T$-dependent part of the
self-energy,  $\delta\Sigma(i\pi T,\xi_0)$,
are small in the parameter $1/E_F\tau$ or $T/E_F$.

\begin{figure}
  \includegraphics[width=0.5\columnwidth]{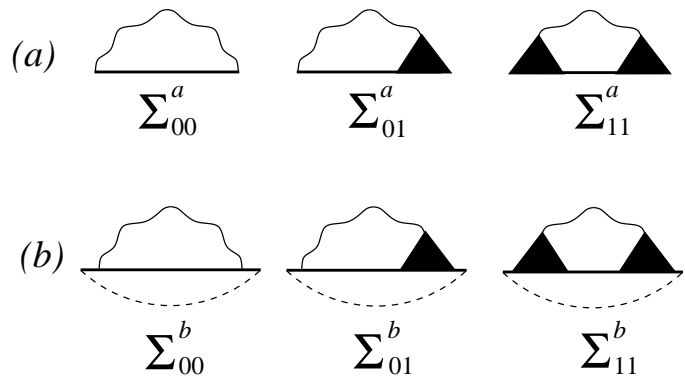}
      \caption{\label{f-sigma}  Self energy diagrams in the first order in the effective interaction (wavy line).
      Black triangles denote impurity ladders $\Gamma$ dressing interaction vertices, Fig.~\ref{f-vertex}.
      (a) ``Simple'' self-energies $\Sigma^a_{ij}$;
      (b) Hikami-box self-energies $\Sigma^b_{ij}$ generated by covering each of  $\Sigma^a_{ij}$ by an impurity line
      (dashed).}
\end{figure}

\begin{figure}
 \includegraphics[width=0.4\columnwidth]{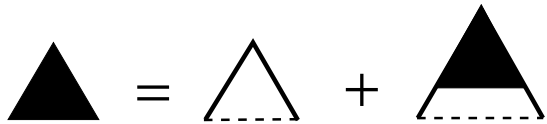}
 \caption{\label{f-vertex} Interaction vertex renormalized by the impurity ladder, $\Gamma(i \omega_{k},{\bf q}).$
 Dashed line represents scattering on impurity.}
\end{figure}

Let us list important elements which are necessary for calculation of the
interaction-induced part of the self-energy (Fig.~\ref{f-sigma}).
Each contribution to the self-energy has exchange and Hartree parts.
We first address the exchange contribution (the Hartree terms can be
written in a similar way).
It contains the angle-averaged Green's function covered by the effective
interaction line. The corresponding vertices may be
dressed by impurity ladders~(Fig.~\ref{f-vertex}). Notice that the renormalized vertex
includes at least one impurity line. Finally, when the interaction line changes the signs
of Matsubara frequencies at vertices, an additional diagram (we term
it a ``Hikami-box diagram'') with a single
impurity line covering the whole block is to be included (Fig.~\ref{f-sigma}b).

We split $\Sigma_{\rm ee}(i\e_n,m\omega_c)$ into three contributions, corresponding
to different possibilities of dressing the two interaction vertices by impurities,
\be
\Sigma_{\rm ee}=\Sigma_{00}+2\Sigma_{01}+\Sigma_{11}.
\label{00+01+11}
\ee
Here the subscripts $i,j=0,1$ indicate whether the corresponding vertex is dressed
by an impurity ladder. The factor 2 in front of $\Sigma_{01}$ term
reflects two possibilities of dressing one of the interaction
vertices.
Each of the terms $\Sigma_{ij}$ is a sum of two contributions,
$\Sigma_{ij}=\Sigma^a_{ij}+\Sigma^b_{ij},$ where $\Sigma^a_{ij}$ is the
``simple'' self-energy (Fig.~\ref{f-sigma}a) and $\Sigma^b_{ij}$ is its Hikami-box counterpart (Fig.~\ref{f-sigma}b).
We note that in Ref.~\onlinecite{MMR} only  one out of six diagrams
(namely, $\Sigma_{01}^a$) was taken into account.

The expression for $\Sigma^a_{00}$ in a finite magnetic field reads
\be
\Sigma_{00}^a(i \e_{n}, m\omega_c)=
-T \sum_{\omega_k}\sum_{L}
  \int \frac{d^{2}q}{\left( 2\pi \right) ^{2}}\, J^2_{|L|}(q R_c) V(i \omega_{k},{\bf q})
   \  G\left(i\varepsilon_{n}-i \omega_{k}, m\omega_c+L\omega_c\right),
  \label{eq:sigma00}
\ee
where
\begin{equation}
G(i\varepsilon_{n}, m\omega_c)=
\frac{1}{i\varepsilon_{n}+\mu-(m+1/2)\omega_c-\Sigma(i\varepsilon_{n},m\omega_c)}
\end{equation}
is the Green's function in Landau levels representation, $V(i \omega_{k},{\bf q})$
is the effective interaction,
and $J^2_{|L|}(q R_c)$ describes
the bare vertex function connecting Landau levels $m$ and $m+L$
in the quasiclassical limit $m,m+L\gg 1$ (for details see,
e.g., Ref.~\onlinecite{drag} and references therein).
In the expressions for
$\Sigma^a_{01}$ and $\Sigma^a_{11}$ this bare vertex function is multiplied
by $\Gamma(i \omega_{k},{\bf q})$ and $\Gamma^2(i \omega_{k},{\bf q})$, respectively,
where $\Gamma$ is the impurity ladder (Fig.\ref{f-vertex}).

In the limit $B\to 0$, the corresponding self-energies depend on the
momentum ${\bf p}$ instead of the Landau level index $m$.
The interaction vertices are dressed only when the Matsubara energies
at the vertices have opposite signs, which restricts the summation over $\omega_k$
in $\Sigma_{01}$ and $\Sigma_{11}$ to the domain
$\e_n(\omega_{k}-\e_n)>0$:
\begin{eqnarray}
\Sigma_{00}^a(i \e_{n}, \xi_p) & = & -T
  \sum_{\omega_k}
  \int \frac{d^{2}q}{\left( 2\pi \right) ^{2}}\, V(i \omega_{k},{\bf q})\
     G( i \varepsilon_{n}-i \omega_{k}, {\bf p-q}) ,
  \label{eq:sigma00B=0}
  \\
\Sigma_{01}^a(i \e_{n}, \xi_p ) & = & -T
  \sum_{ \e_{n} \left( \omega_{k}- \e_{n}\right)>0}
  \int \frac{d^{2}q}{\left( 2\pi \right) ^{2}}\, V(i \omega_{k},{\bf q})\
    \Gamma\left(i \omega_{k},{\bf q}\right) \
  G( i \varepsilon_{n}-i \omega_{k}, {\bf p-q}) ,
  \label{eq:sigma10B=0}
  \\
\Sigma_{11}^a(i \e_{n}, \xi_p ) & = & -T
  \sum_{ \e_{n} \left( \omega_{k}- \e_{n}\right)>0}
  \int \frac{d^{2}q}{\left( 2\pi \right) ^{2}}\, V(i \omega_{k},{\bf q})\
    \Gamma^2\left(i \omega_{k},{\bf q}\right)\
  G( i \varepsilon_{n}-i \omega_{k}, {\bf p-q}),
  \label{eq:sigma11B=0}
  \end{eqnarray}
where
$G\left( i \varepsilon_{n}, {\bf p}\right) =
\left[i \varepsilon_{n}+\mu-\xi_p+i\text{sgn}\varepsilon_{n}/2\tau-
\Sigma_{ee}(i \varepsilon_{n},\xi_p) \right]^{-1}$ with $\xi_p=p^2/2m$
and
the vertex correction (Fig.\ref{f-vertex}) reads
\begin{equation}
  \label{eq:vertex}
  \Gamma\left(i \omega_{k}, \bq\right) =\frac{1}
 {\sqrt{\left( \left| \omega_{k}\right|\tau +1\right)^2+\left( qv_F\tau\right)^2}-1}.
\end{equation}

To calculate the damping factor of the oscillations, 
we need the self-energy taken at the value of $\xi$ which is
determined by the pole of the Green's
function, $\xi=\xi_0,$ see Appendix B.
According to Eq.~(\ref{FT}), the self-energy is further multiplied by $Z$ in the
damping exponent.
This is equivalent to calculating the following integral:
\bea
Z\Sigma^a_{ij}(i\epsilon_{n},\xi_0) &=& { \sgn\e_n\over 2\pi i}\,
{1+\beta_0\over 1+\alpha_0}\int d\xi_k
 G(i\epsilon_{n},k)\Sigma^a_{ij}(i\epsilon_{n},\xi_k)\nonumber \\
&\simeq&
 \sgn\e_n {v_F^*\over 2\pi i} \int dk
 G(i\epsilon_{n},k)\Sigma^a_{ij}(i\epsilon_{n},\xi_k). 
\label{polexi0}
\eea
We note that the $Z$-factor drops out in the product $Z\Sigma$.
Indeed, the Green's function under the interaction line in the
self-energy contains the $Z$-factor in
the numerator so that in the numerator of the product $Z\Sigma$ we get
the factor $Z^2$. 
However, the $Z$-factor is not a gauge invariant quantity and therefore
 should not appear in the expressions for observables, 
in contrast to the FL-renormalized effective mass. 
At this point we should take into account
the FL renormalization of the two interaction vertices in
$\Sigma$. Since we are interested in
the contribution of relatively slow transferred momenta and
frequencies giving rise to the $T$ 
dependence of $B(T)$, $q\ll k_F$ and $\omega_k\ll E_F$, we can set them to zero
when considering the FL vertex renormalization. Then one can apply the Ward identity
for the FL-interaction dressing of the vertices~\cite{Falko}, which amounts to multiplying
each vertex by a factor $1/Z$. These vertex factors in the denominator
of $B(T)$ cancel $Z^2$ in the numerator of $B(T)$.
This implies that one can  simply discard such  renormalizations, setting
$Z=1$ everywhere, when the observable quantities are calculated.
The FL renormalization then amounts to replacement of the bare band mass $m$,
Fermi velocity $v_F,$ and elastic scattering time, $\tau,$  by the
renormalized parameters, 
$m^*,\, v_F^*$ and $\tau^*$, respectively.
In what follows, we will omit the asterisks,  using
the notation $m,\, v_F$ and $\tau$ for the renormalized quantities.

The relevant contributions to the self-energy are calculated in Appendix B.
Combining all the terms together, we have
\bea
\delta\Sigma(i\e_n,\xi_0)&=&-i\ T\sum_{\omega_{m}>\varepsilon_{n}}\int
\frac{d^{2}q}{\left( 2\pi \right) ^{2}}\,
V(i\omega_m,\bq)K(i\omega_m,\bq),\label{sigmaVK}\\
K(i\omega_m,\bq)&=& \frac{[1+\Gamma(i\omega_m,\bq)]^2}{S(i\omega_m,\bq)}
\left[1-\frac{W}{\tau S^2(i\omega_m,\bq)}\right] - \frac{1}{S_0(i\omega_m,\bq)},
\label{Kwq}
\eea
where
\be\label{qtoS}
S(i\omega_m,\bq)=\sqrt{(|\omega_m|+1/\tau)^{2}+v_{F}^{2}q^{2}}
=\sqrt{W^{2}+v_{F}^{2}q^{2}},\qquad  
W=|\omega_m|+1/\tau
\ee
and
\be
S_0(i\omega_m,\bq)=\sqrt{|\omega_m|^{2}+v_{F}^{2}q^{2}}.
\label{S0}
\ee

An important feature of the kernel function $K(i\omega_m,\bq),$ Eq.~(\ref{Kwq}),
is that it is exactly zero for $q=0$ for arbitrary $\omega_m.$ Indeed,
using $S(i\omega_m,q=0)=W$,  $S_0(i\omega_m,q=0)=|\omega_m|$, and
$\Gamma(i\omega_m,q=0)=1/(W\tau-1)$, we get 
\bea
K(i\omega_m,q=0)&=&\frac{1}{W}\left[1+\frac{1}{W\tau-1}\right]^2
\left[1-\frac{1}{W\tau}\right]-\frac{1}{|\omega_m|}\nonumber
\\
&=&\frac{1}{W-1/\tau}-\frac{1}{|\omega_m|}=0.
\label{K(q=0)=0}
\eea
We stress that this equality only holds when all the contributions to
the self-energy are combined together.
This property of the kernel function is characteristic for the gauge-invariant
quantities in the presence of interaction.
Indeed, the interaction at $q=0$ implies the shift of the chemical
potential and hence can be
gauged out~\cite{finkelstein}. Therefore the contribution of small $q$ to the observables
should be suppressed
by vanishing of the corresponding kernel function. The same situation
is well known for the
interaction-induced correction to the conductivity~\cite{ZNA}.

It is worth discussing a peculiarity of the problem of
magnetooscillations with respect to the gauge invariance.
The gauge invariance of the oscillatory part of the observables
is guaranteed by the fact that the thermodynamic potential
is represented by closed loops. Since the characteristic
spatial scale for such a loop is cyclotron radius $R_c$,
the interaction with momenta $q\ll R_c^{-1}$ should not contribute.
So, if we would find that our result does not satisfy this requirement,
it would mean that the diagrammatic treatment is not sufficient and
should be complemented by the infrared cutoff at  $q \sim R_c^{-1}$.
On a more rigorous level this could be done by using the real-space
path integral approach combining the treatment of magnetooscillations
in the presence of long-range disorder~\cite{MWA} with the quantum
kinetic-equation approach to the interaction effects~\cite{ZNA,CA}
and to magnetotransport~\cite{VaAl}. In this context, it is instructive
to recall the calculation of the dephasing length $l_\varphi$,
where the infrared cutoff at $q\sim l_\varphi^{-1}$ or $q\sim R^{-1}$
arises for the problems of weak localization~\cite{AAK} and
Aharonov-Bohm oscillations~\cite{LM} in a ring of radius $R$,
respectively.

Since we find, however, that the kernel $K(i\omega_m,q)$
governing the perturbative self-energy (\ref{sigmaVK}) does satisfy
$K(i\omega_m,q=0)=0,$ the above cutoff is irrelevant and
the perturbative treatment is sufficient
in the considered regime of strongly damped oscillations,
$T\gg \omega_c$. Indeed, the $q$-integral in (\ref{sigmaVK})
is cut off at $q\sim \omega/v_F\sim T/v_F$ which is much larger
than $R_c^{-1}$ under the above condition.

We also emphasize that the kernel $K(i\omega_m,q)$ given by Eq.~(\ref{Kwq})
vanishes in the clean limit
\be
\left.K(i\omega_m,q)\right|_{\tau\to \infty}=0.
\ee
This implies, in particular, that the correction to the effective mass
found in Refs.~\onlinecite{chubukov1,chubukov,DasSarma} from 
the $\e$- and $p$-dependence of the self-energy 
$\Sigma(\e,p)$ of a clean system,
$
\delta m/m^*\sim T/E_F
$
does not show up in the damping of magnetooscillations, in accordance 
with the statement made in Ref.~\onlinecite{chubukov1}.
In general, $\Sigma(\e,p)$ is not an observable (and not gauge invariant) 
quantity, and thus the above correction $\delta m$ should be at least
treated with caution.

\section {Damping of magnetooscillations: short-range interaction}
\label{s3}

In this section we evaluate $\delta\Sigma(i\e_n,\xi_0)$ in the case
of a weak point-like interaction given by
\be
V(i\omega_m,{\bf q})\equiv U_0.
\ee
We are interested in the correction to the self-energy
to the first order in $\nu U_0$, where $\nu=m^*/2\pi$ is
the density of states per spin direction.
We calculate the contributions $\Sigma_{ij}^a$
and  $\Sigma_{ij}^b$ 
starting from Eqs.~(\ref{deltasigma00a}), (\ref{deltasigma01a}),
(\ref{deltasigma11a}), (\ref{deltasigma00b}),
(\ref{deltasigma01b}), and (\ref{deltasigma11b}) derived in Appendix B.
Using the notation $S(i\omega_m,\bq)$ introduced in Eq.~(\ref{qtoS}),
the vertex factor $\Gamma$ can be presented as 
\be
\Gamma(i \omega_{k},\bq) = \frac{1}{S\tau-1}.
\ee
Performing the momentum integration for $0<q<k_F$, we obtain
\bea
\Sigma_{00}^a\left(i \e_n, \xi_0\right)&=&-i\, T\, U_0
  \sum _{\omega_m>\e_n}
  \int_{0}^{k_F} \frac{q dq}{2\pi}
  \left[\frac{1}{S}-\frac{1}{S_0}\right]\nonumber \\
&=&-\frac{i\, T\, U_0}{2\pi\, v_F^2} \sum_{\omega_m>\e_n} \left[
  \sqrt{E_F^2+W^2}- W - \sqrt{E_F^2+\omega_m^2} +\omega_m 
\right]\nonumber \\
&\simeq& -\frac{i\, T\, \nu U_0}{2 E_F\tau} \sum_{\omega_m>\e_n}
\left[\frac{\omega_m}{\sqrt{E_F^2+\omega_m^2}}-1\right],\\
 2\Sigma_{01}^a\left(i \e_n, \xi_0\right) &+&
\Sigma_{11}^a\left(i \e_{n}, \xi_0\right)
=-\frac{i\, T\, U_0}{2\pi\, v_F^2\, \tau}
  \sum _{\omega_m>\e_n}
  \int_{\omega_{m}+1/\tau}^{\sqrt{E_F^2+\omega_m^2}} dS
\frac{(2S-1/\tau)}{(S-1/\tau)^2
    }
\nonumber \\
&=& -\frac{i\, T\, \nu U_0}{2E_F\tau}
  \sum _{\omega_m>\e_n}
    \left(\ln{E_F^2+\omega_m^2\over \omega_m^2}+{1\over \omega_m\tau} \right).
\label{summshort}
\eea
These expressions are valid independently of the value of the
parameter $T\tau$, i.e. they describe both the diffusive and the
ballistic regimes, as well as the crossover between them.
The logarithmic
term in (\ref{summshort}) comes from $\Sigma_{01}^a$,
while the $1/\omega_m$ term originates from $\Sigma_{11}^a$.

The contributions of the Hikami-box diagrams are given by
\bea
\Sigma^{\rm b}_{00}(i \e_n)&=&\frac{i\ T \nu U_0}{2
  E_F\tau}\sum_{\omega_{m}>\varepsilon_{n}}\left(\omega_{m}+\frac{1}{\tau}\right)
\int_{\omega_{m}+\frac{1}{\tau}}^{\sqrt{E_F^2+\omega_m^2}}\frac{dS}{S^{2}}\simeq-\frac{i\
  T \nu U_0}{2
  E_F\tau}\sum_{\omega_{m}>\varepsilon_{n}}
\left[\frac{\omega_m}{\sqrt{E_F^2+\omega_m^2}}-1\right],\\
2\Sigma^{\rm b}_{01}(i \e_n)&+&\Sigma^{\rm b}_{11}(i \e_n)
=\frac{i\, T\, U_{0}}{2\pi\,
  v_{F}^{2}\tau^{2}}\sum_{\omega_{m}>\varepsilon_{n}}
\left(\omega_{m}+\frac{1}{\tau}\right)
\int_{\omega_{m}+\frac{1}{\tau}}^{\sqrt{E_F^2+\omega_m^2}}
\frac{dS}{S^{2}}\frac{(2S-1/\tau)}{(S-1/\tau)^{2}}
=\frac{i\, T\, \nu U_0}{2E_F\tau}
  \sum _{\omega_m>\e_n} {1\over \omega_m\tau}.
\eea

We see that the Hikami-box contribution exactly cancels the second
($1/\omega_m$) term in (\ref{summshort}).
Thus the total correction to the self-energy reads
\be
\delta\Sigma(i\e_n,\xi_0)=-\frac{i\, T\, \nu U_0}{E_F\tau}
  \sum_{\omega_m>\e_n}\left[\left(\frac{\omega_m}{\sqrt{E_F^2+\omega_m^2}}-1\right)+
    \ln{ \sqrt{E_F^2+\omega_m^2}\over \omega_m }\right].
    \label{summshort-tot}
\ee
The upper limit of the summation over $\omega_m$ is effectively given by $m_{\rm max}\sim
E_F/T$.
The term $[\omega_m/\sqrt{E_F^2+\omega_m^2}-1]$ in
Eq.~(\ref{summshort-tot}),  which originates from
$\Sigma_{00}$, yields
after the summation over $\omega_m>\pi T$  a contribution $\propto T$
in addition to a large $T$-independent contribution, renormalizing $\tau$.
As we will see below, a term $\propto T\ln (E_F/T)$ will arise from the contributions of
$\Sigma_{01}$ and $\Sigma_{11}$, and hence $\Sigma_{00}$ yields only a subleading
contribution to the damping.

To calculate the sum of logarithms in (\ref{summshort-tot}) we write
\be
 \sum_{\omega_m>\e_n} \ln{ \sqrt{E_F^2+\omega_m^2}\over \omega_m} =  \sum_{\omega_m>\e_n}
   \left[\ln{E_F\over \omega_m} +  \ln{ \sqrt{E_F^2+\omega_m^2}\over E_F}\right].
\ee
For  $T\ll E_F$ the second sum can be replaced by the integral,
yielding a $T$-independent contribution.
We further use the identity
\be
\sum_{m=1}^{M}\ln {N\over m}=M\ln N - \ln \Gamma(M+1), \nonumber
\ee
and the Stirling's formula
$$\ln\Gamma(M+1)= M\ln M-M+{1\over 2}\ln(2\pi M) + ...$$
for $M\gg 1$, where $\Gamma(x)$ is the gamma-function.
As a result, we get for arbitrary positive $\e_n$
\be
 \delta\Sigma(i\e_n,\xi_0)\simeq -\frac{i\, T\, \nu U_0}{E_F\tau} \
  \left\{ {c_1\ E_F\over T}
-\frac{\e_n}{2\pi T}\ln{c_2E_F\over T}+\ln\left[{1\over
    \sqrt{2\pi}}\Gamma\left(\frac{\e_n}{2\pi T}+{1\over
      2}\right)\right]\right\},
\label{delta-sigma-vert-short-result}
\ee
where $c_{1,2}$ are constants of order unity.
We see that the contribution to $\delta\Sigma(i\pi T,\xi_0)$ containing
vertex corrections to the interaction line
has an additional factor
$\ln(E_F/T)$ as compared to the $T$-dependent part of $\delta\Sigma_{00}$.
In Eq.~(\ref{delta-sigma-vert-short-result}), we have 
absorbed the contribution of $\delta\Sigma_{00}$ into the upper
cutoff of the log-term which is given by $E_F$ up to a factor of order unity.
Furthermore, the same can be done with the 
last term in Eq.~(\ref{delta-sigma-vert-short-result}
which at $\e_n=\pi T$ also yields a linear-in-$T$ contribution. 
Equation (\ref{delta-sigma-vert-short-result}) thus
translates for $T\gg \omega_c$ into the following expression for the damping
exponent $B(T)=-2\pi i \delta\Sigma(i\pi T,\xi_0)/\omega_c$:
\begin{equation}
 B(T)=-c_1\, \nu U_0 {\pi \over \omega_c\tau}
 +
\frac{\pi T}{\omega_c}\
{\nu U_0\over E_F \tau}\ln\frac{c_2 E_F}{ T}.
\label{FT-short-result}
\end{equation}
The first term in Eq.~(\ref{FT-short-result}) describes the $T$-independent
FL-renormalization of $\tau$ due to vertex corrections and should be included in the
effective relaxation time $\tau^*$, as was mentioned after
Eq.~(\ref{taustar}) in Sec.~\ref{s2}.
The second term represents the  $T$-dependent contribution to the
damping factor that we are interested in and is analyzed in the next section.

\section {Interpretation: effective mass vs quantum scattering time}
\label{s4}

The above result (\ref{FT-short-result}) can be interpreted in terms
of corrections to the effective mass (or $\omega_c$) and the elastic scattering rate
entering the standard formula (\ref{A10}). These corrections
come from the interplay of disorder and interaction. Writing
\begin{eqnarray}
A_1(T)&=&{4\pi^2 T\over \omega_c}\exp\left[-{2\pi^2 T\over(\omega_c+\delta\omega_c)}
- {\pi\over (\omega_c+\delta\omega_c)}\ {1\over(\tau+\delta\tau)}\right]\nonumber \\
&\simeq&A_1^{(0)}(T)\exp\left[{2\pi^2
    T\over\omega_c}{\delta\omega_c\over\omega_c}\right]
\exp\left[{\pi\over \omega_c\tau}\left({\delta\omega_c\over\omega_c}+
{\delta\tau\over\tau}\right)\right]\nonumber \\
&\simeq&A_1^{(0)}(T)\exp\left[-{2\pi^2 T\over\omega_c}{\delta m\over m}\right]
\exp\left[{\pi\over \omega_c\tau}\left(-{\delta m\over m}+
{\delta\tau\over\tau}\right)\right],
\label{Acorrected}
\end{eqnarray}
we conclude that
\begin{equation}
B(T)=-{2\pi^2 T\over\omega_c}{\delta m\over m}-{\pi\over
  \omega_c\tau}\left({\delta m\over m}-
{\delta\tau\over\tau}\right).
\label{FTviamtau}
\end{equation}
It is worth noting that the FL-renormalization does not affect the
product $\omega_cm=e B$.

Comparing (\ref{FT-short-result}) and  (\ref{FTviamtau}) [we recall that the
first term in Eq.~(\ref{FT-short-result}) is absorbed in $\tau$],
we see that
the $T\ln T$ dependence of the damping factor
could in principle originate either from  the $\ln T$ correction
to the effective mass, or from the $T\ln T$-type correction to $\tau$.
This led the authors of Ref.~\onlinecite{MMR} to the conclusion that
the nonlinear $T-$dependence of
the damping factor may be equivalently interpreted either as a
$T-$dependent renormalization of the
effective mass or as a $T-$dependent Dingle temperature.
It is clear, however, that these two possibilities correspond to different
physical processes.

\subsection{Self-energy at real energies: analytical continuation}
\label{s4.1}

To identify the physical origin of the leading contribution to the damping
it is instructive to obtain $B(T)$ using the expression for
the self-energy analytically continued to real values
of energies $\e_n\to -i\e$ .
Performing the analytical continuation
to real energies $\e$
and real frequencies $\omega$ in Eq.~(\ref{sigmaVK}),
we get
\be
\delta\Sigma(\e,\xi_0)=\frac{\nu U_0}{E_F\tau}\int_{-\infty}^\infty
{d\omega\over 4\pi}\tanh\left( \frac{\e-\omega}{2
    T}\right)\left\{\ln\left(\frac{\sqrt{E_F^2-\omega^2}}{-i\omega}\right)
+\left(\frac{-i\omega}{\sqrt{E_F^2-\omega^2}}-1\right)\right\},
\label{dssr-ac}
\ee
so that the real part of the self-energy is given by
\bea
\re\delta\Sigma(\varepsilon,\xi_0)&=&\frac{\nu U_0}{4\pi E_F\tau}\left\{
\int_{0}^{E_F} d\omega \left[\tanh\left( \frac{\e-\omega}{2
      T}\right)+\tanh\left( \frac{\e+\omega}{2 T}\right)\right]
\left[\ln\left(\frac{\sqrt{E_F^2-\omega^2}}{\omega}\right)-1\right]\right.
\nonumber \\
&+& \left.
\int_{E_F}^{\infty}d\omega\left[\tanh\left( \frac{\e-\omega}{2
      T}\right)+\tanh\left( \frac{\e+\omega}{2 T}\right)\right]
\left[
  \ln\left(\frac{\sqrt{\omega^2-E_F^2}}{\omega}\right)
+\frac{\omega}{\sqrt{\omega^2-E_F^2}}-1\right]
\right\}
\nonumber \\
 &\simeq &  \varepsilon \, \frac{\nu U_0}{2 \pi E_F\tau} \ln \frac{E_F}{{\rm
     max}[|\e|,T]}.
\label{re-delta-sigma-short}
\eea
The leading contribution here comes from the term $\ln(E_F/\omega)$ in
the first integral over $\omega<E_F$
while other terms only rescale the ultraviolet cut-off $E_F$ of
  the logarirthm by a constant of order unity, which
is beyond the accuracy of our quasiclassical approximation.

The imaginary part of $\delta\Sigma$ reads
\bea
\im\delta\Sigma(\varepsilon,\xi_0)&=&\frac{\nu U_0}{4\pi E_F\tau}
\int_{0}^{E_F} d\omega \left[\tanh\left( \frac{\e-\omega}{2
      T}\right)-\tanh\left( \frac{\e+\omega}{2 T}\right)\right]
\left[\frac{\pi}{2}-\frac{\omega}{\sqrt{E_F^2-\omega^2}}\right]
\nonumber \\
 &\simeq & -{\rm const}\frac{\nu U_0}{\tau}
  +\frac{\nu U_0}{2 E_F\tau}\, T\, \ln\left[2 \cosh\left({\e \over
        2T}\right)\right],
\label{im-delta-sigma-short}
\eea
where the $T$-dependent term has the following asymptotics:
\begin{equation}
  T \,\ln[2\cosh(\e/2T)]=
 \left\{\begin{array}{ll}
   \varepsilon/2, \quad & \e\gg T,\\[0.2cm]
 T \ln 2, \quad & \e\ll T.
  \end{array}\right.
  \label{deltatau-asym}
\end{equation}
The contribution of the term $\omega/\sqrt{E_F^2-\omega^2}$ to the
integral in (\ref{im-delta-sigma-short}) is $T$-independent up to
small corrections of order of $\nu U_0(T/E_F)^2/\tau$ which are beyond the accuracy of the
calculation.

Having calculated $\re\Sigma$ and $\im\Sigma$ for real energies $\e$, we can
determine
$\delta m$ and $\delta \tau$. Indeed, the magnitude of the first harmonics of
the magnetooscillations of the thermodynamic density of states
is expressed through the real-$\e$
self-energy $\delta\Sigma(\e)$
as follows:
\bea
A_1(T) &=& {4\pi^2 T\over \omega_c} 
\int d\e \ \left[-{\partial n_F(\e)\over \partial \e}\right]
A_1(\epsilon,T),\label{a1T}\\
A_1(\epsilon,T)&=&
\exp\left\{\frac{2\pi
    i}{\omega_c}[\varepsilon-\re\delta\Sigma(\varepsilon,\xi_0)]\right\}
\exp\left\{-\frac{\pi}{\omega_c\tau}
+\frac{2\pi}{\omega_c}\im\delta\Sigma(\varepsilon,\xi_0)\right\},
\label{A1eT}
\eea
where $n_F(\e)=[1+\exp(\e/T)]^{-1}$ is the Fermi distribution function.

In analogy with Eq.~(\ref{Acorrected}) we represent the energy-dependent
amplitude $A_1(\epsilon,T)$ in terms of energy- and temperature-dependent
corrections to the quantum scattering time and mass,
$\delta\tau(\e, T)$ and $m(\e,T)$:
\be
A_1(\epsilon,T)=\exp\left\{{2\pi i \e \over\omega_c}\left[1+{\delta m(\e,T)\over m}\right]\right\}
\exp\left\{-{\pi\over \omega_c\tau}\left[1+{\delta m(\e,T)\over m}-
{\delta\tau(\e,T)\over\tau}\right]\right\}.
\label{A1eTtm}
\ee
Comparing (\ref{A1eTtm}) with (\ref{A1eT}), we express $\delta\tau(\e, T)$ and
$m(\e,T)$
through $\re \Sigma(\e)$ and $\im \Sigma(\e)$ as follows:
\bea
{\delta m(\e,T)\over m}&=& -\frac{\re \delta\Sigma(\e,T)}{\e},
\label{deltamass}\\
{\delta\tau(\e,T)\over\tau}&=&  2 \tau \im \delta\Sigma(\e,T)  +
{\delta m(\e,T)\over m}.
\label{deltatautau}
\eea
Using (\ref{re-delta-sigma-short}) and (\ref{im-delta-sigma-short}) in
combination with (\ref{deltamass}) and (\ref{deltatautau}),
we obtain
\bea
{\delta m(\e,T)\over m}&=& -  \frac{\nu U_0}{2 \pi E_F\tau} \ln \frac{E_F}{{\rm
     max}[|\e|,T]},
\label{deltamasseT-result}\\
{\delta\tau(\e,T)\over\tau}&=&  \nu U_0 \frac{T}{ E_F} \ln\left[2 \cosh\left({\e \over
        2T}\right)\right] - \frac{\nu U_0}{2 \pi E_F\tau} \ln \frac{E_F}{{\rm
     max}[|\e|,T]}.
\label{deltataueT-result}
\eea

The integration in (\ref{a1T}) sets in effect  $\epsilon\sim T$ in the above
expressions. The $T$-dependent corrections to the effective mass and the quantum
scattering time extracted experimentally with the help of
Eq.~(\ref{Acorrected}) are thus given by
\bea
{\delta m(T)\over m}&=& -  \frac{\nu U_0}{2 \pi E_F\tau} \ln \frac{E_F}{T},
\label{deltamassT-result}\\
{\delta\tau(T)\over\tau}
&=& \nu U_0 \frac{T}{ E_F} - \frac{\nu U_0}{2 \pi E_F\tau} \ln \frac{E_F}{T}.
\label{deltatauT-result}
\eea
In (\ref{deltatauT-result}) we assume 
that $\pi \nu U_0 T/E_F\ll \omega_c\tau,$
expand $\exp[\pi \delta\tau(\e,T)/\omega_c\tau^2],$ and then average the term
$\ln[2 \cosh(\e/2T)]$ [which is a real-energy counterpart of the last term in 
Eq.~(\ref{delta-sigma-vert-short-result})] 
with $-\partial n_F(\e)/\partial \e$.

It is clear from these results that
the leading term in $B(T)$ [proportional to $T \ln(E_F/T)$,
Eq.~(\ref{FT-short-result})] originates from the real part of the
self-energy, Eq.~(\ref{re-delta-sigma-short}),
i.e. from renormalization of the effective mass,
which affects incommensurability
of the oscillations at different values of energy $\varepsilon$.
The contribution to $B(T)$ of the imaginary part of the self-energy 
[corresponding to the last term in Eq.~(\ref{delta-sigma-vert-short-result})], 
which is governed in the ballistic regime by the renormalization of the
scattering time, is smaller by a factor $\ln(E_F/T)$. 
In the expression for the damping,  Eq.~(\ref{FT-short-result}), 
this contribution is absorbed in the numerical constant $c_2$ in 
the upper cutoff of the logarithm.

The obtained result for the interaction-induced correction to the
scattering time $\tau$, Eq.~(\ref{deltatauT-result}), agrees, up to a
factor ${1\over 2}$,  with the
correction to the transport time following from the calculation of
conductivity correction in the ballistic regime
in Ref.~\onlinecite{ZNA}. This is exactly what one
would expect on physical grounds. Indeed, it is known that the
conductivity correction \cite{ZNA} can be understood as governed by an
additional, predominantly back-scattering, contribution to the
scattering cross-section related to the dressing of an impurity by
Friedel oscillations. Since this contribution is concentrated near the
scattering angle $\phi=\pi$, the correction to the momentum relaxation rate is
larger by the factor $1-\cos\phi\simeq 2$ than the correction to the
total scattering rate. In Sec.~\ref{s3.2} we will corroborate the
results of this subsection by  an explicit
calculation of the contribution to the impurity scattering rate due to
Friedel oscillations.

Up to now we calculated the exchange contribution to the self-energy.
For the point-like interaction, the Hartree term has opposite sign and
is twice larger in magnitude than the exchange term
due to the spin summation. This simply reverses the sign of the corrections
to the damping factor.

\subsection{Calculation of $\delta\tau$ from the scattering off Friedel oscillations.}
\label{s3.2}

In this subsection we calculate the correction to the total elastic
scattering time $\delta\tau$ in a different, physically more transparent way,
considering the scattering off impurities dressed by Friedel
oscillations~\cite{rudin,ZNA}. We will demonstrate how the result
(\ref{deltatauT-result}) is reproduced in this way. In particular,
this will confirm once more
that there is no $T\ln T$ term in $\delta\tau$ and therefore
the leading $T\ln T$ contribution to the damping factor comes from $\delta m$.

\begin{figure}
  \includegraphics[width=0.3\columnwidth]{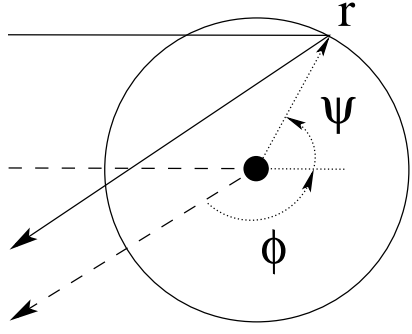}
  \vskip0.5cm
    \caption{\label{friedel}  Schematic illustration of the scattering off Friedel oscillations.
    The black dot in the middle represents a short-range impurity which creates the oscillatory correction to the electron
    density around it. The circle represents the equipotential line of the effective impurity potential.
    The correction to the impurity cross-section at the angle $\phi$ arises due to the interference of
    two electronic waves~\cite{rudin,ZNA},
    one of which (dashed line) scattered by the impurity and another (solid line)
    by the Friedel oscillations at a point parametrized by the distance $r$ from the impurity and the angle $\psi.$}
\end{figure}

We start with the expression relating the total elastic
scattering rate and the scattering cross-section $S(\phi)$ of a single impurity,
\begin{equation}
{1\over \tau_q(\e,T)}=n_{\rm imp} v_F \int {d\phi\over 2\pi} S(\phi),
\label{tauq}
\end{equation}
where $n_{\rm imp}$ is the concentration of impurities.
The expression for the transport scattering time $\tau_{\rm tr}$ determining
the conductivity differs from (\ref{tauq}) by a factor $1-\cos\phi$ in the integrand:
\begin{equation}
{1\over \tau_{\rm tr}(\e,T)}=n_{\rm imp} v_F \int {d\phi\over 2\pi} S(\phi)(1-\cos\phi).
\label{tautr}
\end{equation}
We note that the two times, $\tau_q$ and $\tau_{\rm tr},$ though equal
for the point-like impurities in the non-interacting case
[for which $S(\phi)=(n_{\rm imp} v_F\tau)^{-1}={\rm const}(\phi)$], differ from each other
when the scattering off Friedel oscillations is taken into account.

The impurity scattering cross-section for dressed impurities reads
\be
\label{S-phi}
S(\phi)=\frac{2\pi \nu}{v_F} \left|V\left(2k \sin{\phi\over 2}\right)\right|^2
\simeq S_0 + \frac{2\pi \nu V_0}{v_F} \, 2\re \delta V\left(2k \sin{\phi\over 2}\right).
\ee
Here $S_0=2\pi \nu V_0^2/v_F$ is the bare impurity scattering cross-section and
$V_0=\int d^2 r V_0(r),$ where $V_0(r)$ is the bare point-like impurity potential.
The cross-section $S(\phi)$ depends on energy $\epsilon$ of an electron through
$k=k_F+\epsilon/v_F$ in the Fourier transform
$V(q)=V_0+\delta V(q)$ of the effective impurity
potential,
renormalized by the Friedel oscillations of the electron density. For $r\gg k_F^{-1}$ the oscillatory
correction to the electron density reads
\begin{equation}
\label{delta-rho}
\delta\rho(r)=-\nu V_0 \frac{(2 \pi r T/v_F)}{\sinh(2\pi r T/v_F)}\
\frac{\sin 2k_F r}{\pi r^2}.
\end{equation}

The correction to the impurity scattering
potential $\delta V(r)$ due to scattering off the Friedel oscillations
is proportional, for the short-range interaction, to the
electron density at point $\br$.
Similarly to the consideration of Sec.~\ref{s3} and Sec.~\ref{s4.1},
we will concentrate on the exchange part of this correction,
\be
\delta V(r)=-\frac{1}{2}U_0 \delta\rho(r).
\ee
To calculate the correction to the impurity cross-section,
we need the Fourier transform of $\delta V(r),$
\begin{equation}
\label{deltaV}
\delta V\left(2k \sin{\phi\over2}\right) =  \frac{U_0}{2} \int d^2r \,
\delta\rho(r)\, \exp\left(2 i\ k r \sin{\phi\over2} \cos\psi\right),
\end{equation}
where $\psi$ is the polar angle of ${\bf r}$, see Fig.~\ref{friedel}.
Substituting (\ref{delta-rho}) and (\ref{deltaV}) into (\ref{S-phi}),
we find the interaction-induced correction to the scattering cross-section,
\be
\delta S(\phi)= \nu U_0\, S_0\,  \int {d\psi\over 2\pi} \int  dr
\frac{(2\pi T/v_F)}{ \sinh(2\pi r T/v_F)}
\left\{ \sin\left(2r\left[k_F+k\cos\psi\sin{\phi\over 2}\right]\right)
+\sin\left(2r\left[k_F-k\cos\psi\sin{\phi\over 2}\right]\right)\right\}.
\label{delta-cross-section}
\ee
Performing the integration over $r$ we get
\begin{equation}
\delta S(\phi)= S_0\, {\nu U_0 \over 4}\, \int
d\psi\left[\tanh\beta_{+}+\tanh\beta_{-}\right],
\label{deltaS}
\end{equation}
where
\begin{equation}
\beta_{\pm}={v_F\over 2 T}\left(k_F\pm k\cos\psi\sin{\phi\over 2}\right)={E_F\over T}
\left[1\pm \left(1+{\e\over 2 E_F}\right) \cos\psi\sin{\phi\over 2}\right].
\end{equation}

\begin{figure}
  \includegraphics[width=0.4\columnwidth]{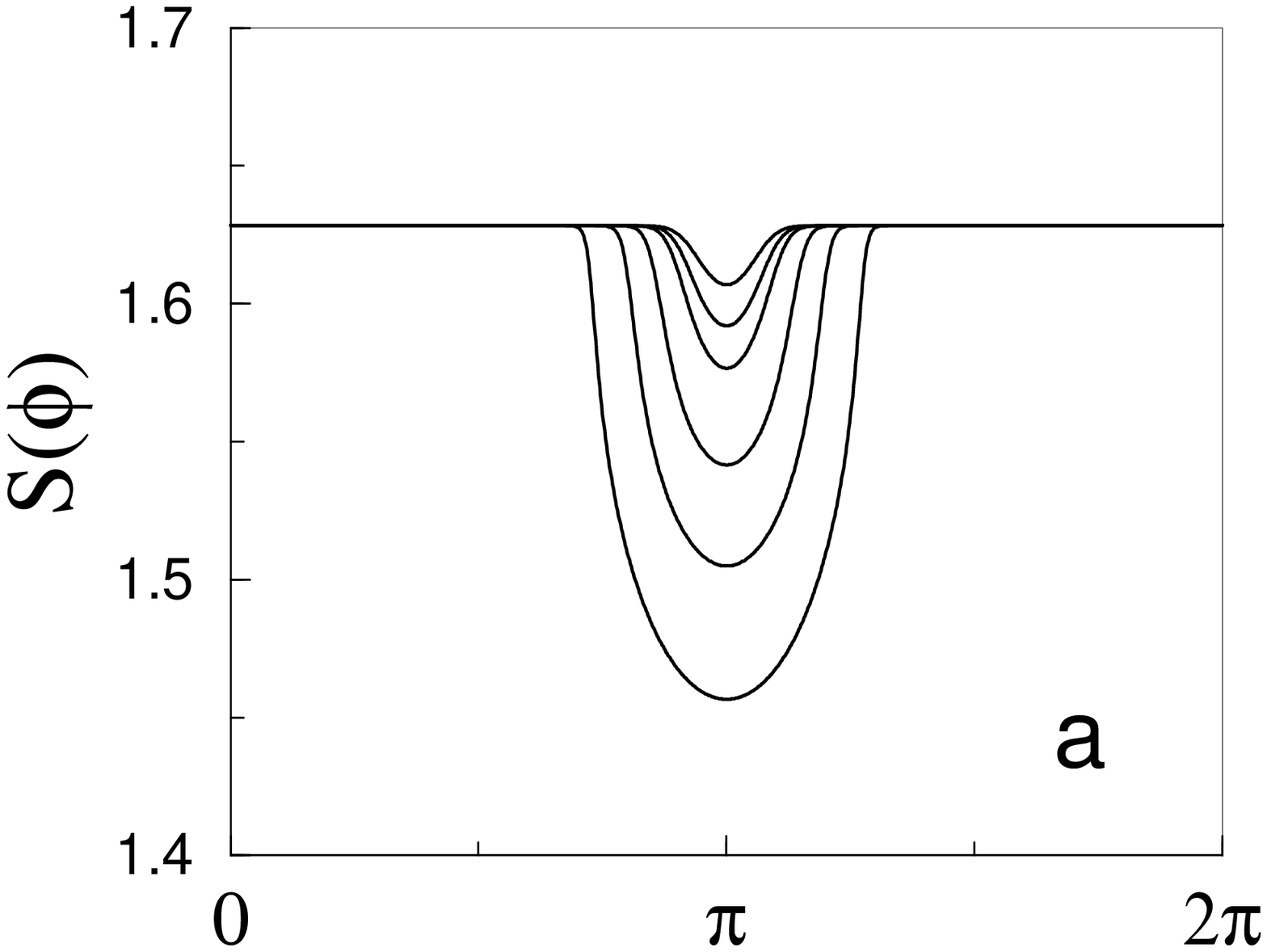}\hspace*{0.5cm}
  \includegraphics[width=0.4\columnwidth]{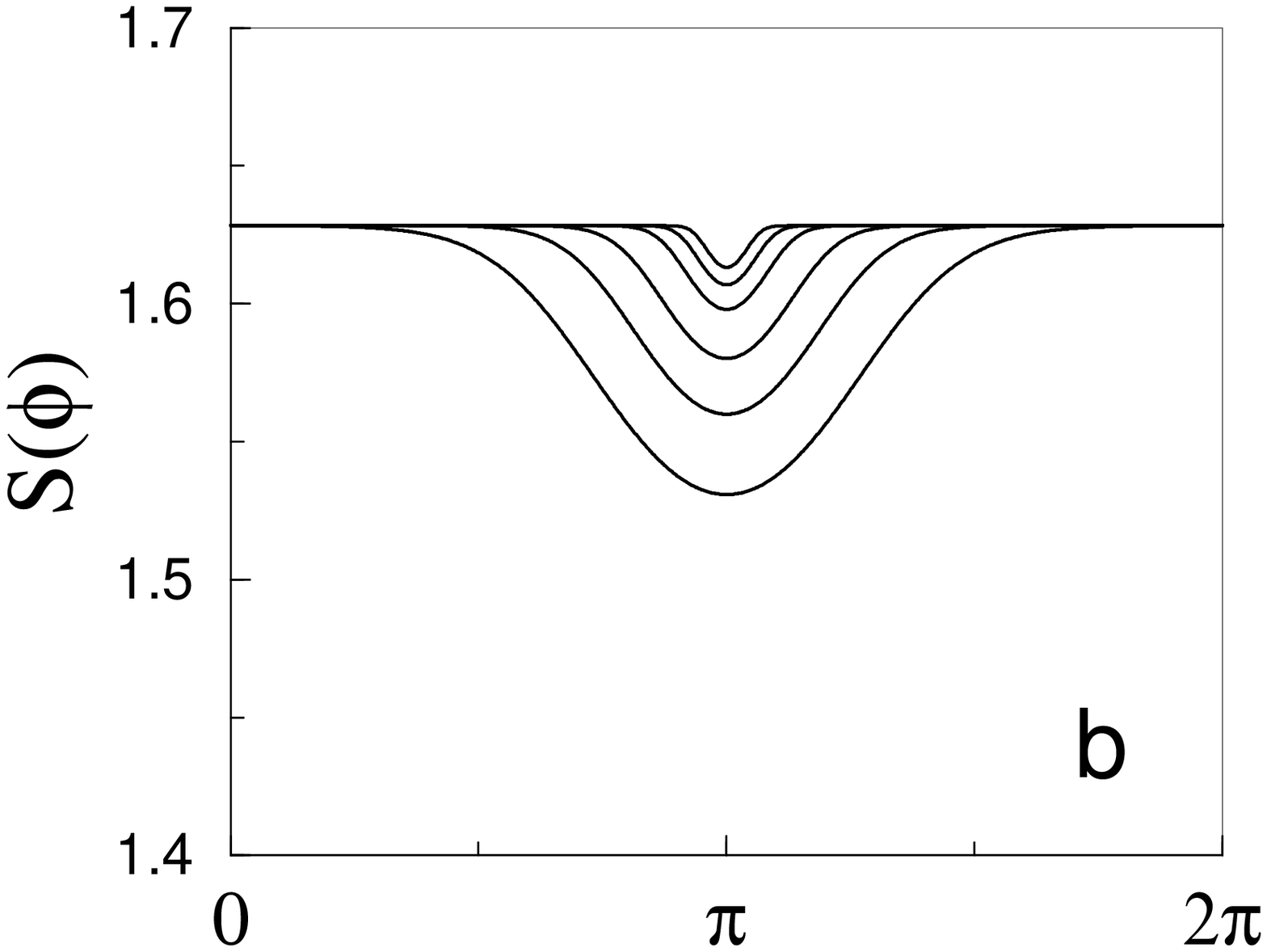}
  \vskip0.5cm
    \caption{\label{cross-section}  Differential impurity cross-section $S(\phi)$
    (in units of the bare cross-section $S_0$)
    renormalized by the scattering off Friedel oscillations
    (with only the exchange contribution taken into account) calculated for $\nu U_0=0.2$ and
    (a) fixed $T=0.01 E_F$ and
    $\epsilon/E_F=0,\ 0.01,\ 0.02,\ 0.05,\ 0.1,\ 0.2,$ from top to bottom;
    (b) fixed $\epsilon=0$ and $T/E_F=0.005,\ 0.01,\ 0.02,\ 0.05,\ 0.1,\ 0.2,$ from top to bottom.}
\end{figure}

For $\beta_{\pm}\gg 1,$ i.e. for most values of the scattering angle $\phi$
except for those corresponding to the backscattering ($\phi\approx \pi$),
we see that the scattering cross-section does not depend on $\phi$
up to exponentially small corrections of order
${\cal O}(\exp[-E_F/T])$:
\begin{equation}
S(\phi)=S_0\left[1+\pi \nu U_0\right], \qquad |\phi-\pi|\gg ({\rm max}[T,\e]/E_F)^{1/2}.
\label{S-flat}
\end{equation}
On the other hand, in the vicinity of $\phi=\pi$  the cross-section has
a  ``hump'', see Fig.~\ref{cross-section}, of the width and the height scaling as
$\delta\phi,\ \delta S(\phi)/S_0 \sim ({\rm max}[T,\e]/E_F)^{1/2}$.
The explicit expression for $S(\phi)$ at $T=0$ can be found in Ref.~\onlinecite{ZNA}.
In Fig.~\ref{cross-section} we plotted $S(\phi)$ for fixed $T$ and several values of $\e$ (Fig.~\ref{cross-section}a)
and for fixed $\e=0$ and several values of $T$ (Fig.~\ref{cross-section}b).

To calculate the $(\e,T)$-dependent correction to $1/\tau_q$ (which is determined by the ``hump''),
we expand $\sin(\phi/2)$ around $\phi=\pi.$ Furthermore, we expand $\cos\psi$ around $\psi=\pi$ and $\psi=0$
in the expressions for $\beta_+$ and $\beta_{-}$, respectively.
This corresponds to the two possible interfering paths propagating along the horizontal line in Fig.~\ref{friedel}
with the scattering off the Friedel oscillation occurring either to the left or to the right from the impurity.
Denoting $\delta\psi=x$, $\delta\phi/2=y$, $z^2=x^2+y^2,$ and $\omega=z^2 E_F,$ we get from Eq.~(\ref{deltaS})
\bea
\int {d\phi\over 2\pi} \delta S(\phi)&\simeq& S_0\, {\nu U_0 \over 2\pi }\, \int dx \int dy \,
\tanh\left\{\frac{E_F}{T}\left[1- \left(1+\frac{\epsilon}{2E_F}\right)
\left(1-\frac{x^2+y^2}{2}\right)\right]\right\}
\nonumber \\
&\simeq& S_0\, {\nu U_0} \int z dz \, \tanh\left\{\frac{E_F}{T}\left[\frac{z^2}{2}-\frac{\e}{2E_F}\right]\right\}
=S_0\, \frac{\nu U_0}{2E_F} \int_0^{\sim E_F} d\omega\, \tanh\left(\frac{\omega-\e}{2T}\right)
\label{deltaSintw}
\eea
Using (\ref{tauq}) and (\ref{deltaSintw}), we
find the total quantum scattering rate,
\begin{eqnarray}
{1\over \tau_q(\e,T)}&=&{1\over \tau}+\delta\left({1\over \tau_q}\right),\\
\delta\left({1\over \tau_q}\right)&=&
-{\delta\tau_q(\e,T)\over\tau^2} =
{\nu U_0\over \tau} \left[{\rm const} -
\frac{  T}{ E_F}\ln\left(2\cosh\frac{\e}{2T}\right)-\frac{\e}{2E_F}\right].
\label{deltatauq-friedel}
\end{eqnarray}
This result for the interaction-induced correction to $\tau_q$
agrees (to the leading order in $1/T\tau$ corresponding to the
ballistic regime) with that obtained from the imaginary part of the
self-energy,  Eqs.~(\ref{deltataueT-result}) and
(\ref{deltatauT-result}). More accurately, 
Eq.~(\ref{deltatauq-friedel}) differs from Eq.~(\ref{deltataueT-result})
by the last ($T$-independent) term $\nu U_0 \e/2E_F\tau$,  
which drops out after the thermal averaging with
$-\partial n_F(\e)/\partial \e$ and thus does not contribute to
$\delta\tau_q(T)$, Eq.~(\ref{deltatauT-result}). This term is in a
sense anomalous, since it arises from the ultraviolet limit of the
$\omega$-integration in Eq.~(\ref{deltaSintw}). In fact, one could
question the validity of this contribution, since we used the
asymptotic, large-$r$ form of the Friedel oscillations in
Eq.~(\ref{delta-rho}). One can check, however, that the same result
[up to an irrelevant additive constant independent of $T$ and $\e$: in Eq.~(\ref{S-flat}) $\pi \nu U_0$
is replaced by $2\nu U_0$] is
obtained from a calculation using the exact form of the Friedel
oscillations \cite{aristov}. The appearance of this linear-in-$\e$
term is related to the violation of the particle-hole symmetry in the
parabolic spectrum; this term did not appear in the diagrammatic
calculation of Sec.~\ref{s4} where the spectrum was linearized. 
What however enters the experimental damping of
magnetooscillations is $1/\tau_q$ integrated over the energy with an
even function $-\partial n_F(\e)/\partial \e$. Therefore, we are in
fact interested in the collision rate symmetrized with respect to 
$\e\to -\e$. Performing this symmetrization in
Eq.~(\ref{delta-cross-section}), we get a result determined solely by
the infrared scale, $r\sim v_F/T$, yielding Eq.~(\ref{deltatauq-friedel})
without the last, linear-in-$\e$ term.

Finally, using (\ref{tautr}), we see that the correction to
the transport rate is larger than $\delta(1/\tau_q)$ by a factor of 2.
The corresponding correction to the conductivity reproduces the result
of Ref.~\onlinecite{ZNA} in the ballistic limit.

\section {Damping of magnetooscillations: Coulomb interaction}
\label{s5}
\begin{figure}
  \includegraphics[width=0.5\columnwidth]{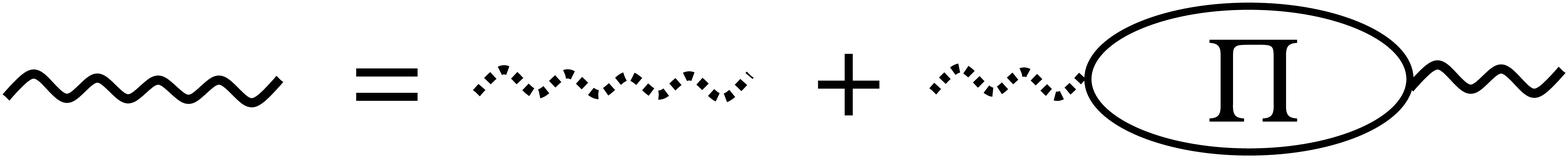}
  \caption{\label{f-line} Diagrammatic equation for the effective interaction
  line (bold wavy line) in the random-phase approximation, Eqs.~(\ref{Vrho}) and (\ref{Vsigma}).
  Dotted wavy line represents the bare interaction,
  $V_0(q)+F_0^\rho/2\nu$ in the singlet channel or $F_0^\sigma/2\nu$ in the triplet channel.
  The bubble $\Pi$ is the polarization operator, Eq.~(\ref{P}).
  }
\end{figure}

We turn now to the Coulomb interaction.
In the case of Coulomb interaction, one should take into account the
dynamical screening of the interaction within the random phase approximation (RPA), see Fig.~\ref{f-line}.
In what follows we use for simplicity the so-called $F_0$-approximation~\cite{ZNA}, which
retains only the zeroth harmonics of the Fermi-liquid
constants $F^{\rho}_{0}$ and $F^{\sigma}_{0}$ in the charge and spin
channels, respectively.
Then the effective interaction propagator in the
charge channel (combining the exchange term and the singlet
contribution of the Hartree-type
interaction) reads \cite{ZNA,GM}
\be
V^\rho\left(i \omega _{m},q\right) =\left[(V_{0}(q) +F_0^{\rho
  }/2\nu)^{-1}+{\Pi}\left(i \omega _{m},q\right)\right]^{-1},
\label{Vrho}
\ee
where ${\Pi}\left(i \omega _{m},q\right)$ is the polarization operator
\be
{\Pi}\left(i \omega _{m},q\right) =2\nu \left[ 1-\left|
\omega_{m}\right| \tau \Gamma\left(i \omega _{m},q\right) \right],
 \label{P}
\ee
$\Gamma\left(i \omega _{m},q\right)$ is the impurity ladder,
and $\nu=m/2\pi$ is the density of states per spin direction.
The triplet contribution to the effective interaction arises from the ladder of Hartree-type
interaction blocks and
reads
\be
V^{\sigma}\left(i \omega _{m},q\right) =\left[2\nu /F_0^{\sigma}+%
{\Pi}\left(i \omega _{m},q\right)\right]^{-1}.
\label{Vsigma}
\ee

\subsection{Singlet channel}
\label{s5.1}

The main difference as compared to the case of the
short-range weak interaction considered in Sec. III is the nontrivial form
of the dynamically screened Coulomb interaction
(one should take into account the renormalization of the interaction
by polarization operator).
Using
$$ V_{0}(q)=\frac{2\pi e^{2}}{q}$$
and neglecting $F_{\rho }^{0}$, we get
\begin{eqnarray}
V^{\rho}(i\omega _{m},q)&=&\frac{V_{0}(q)}{1+V_{0}(q)\Pi (i\omega _{m},q)}
= \frac{2\pi e^{2}/q}{1+(2\pi e^{2}/q) 2\nu [1-|\omega _{m}|\tau
  \Gamma (i\omega _{m},q)] }
\nonumber \\[0.5cm]
&=&
{1\over 2\nu} \frac{\kappa(S-1/\tau)}{(q+\kappa)(S-1/\tau)-\kappa (W-1/\tau)}.
\label{Uscreened-general}
\end{eqnarray}
Here we use the standard notation $\kappa =4\pi \nu e^{2}$
for the inverse Thomas--Fermi screening radius and use the short-hand notation $S$ for
$
S(i\omega_m,\bq)\equiv\sqrt{W^{2}+v_{F}^{2}q^{2}}
$
with $W=|\omega_m|+1/\tau$.

For $q\ll \kappa$, neglecting $q$ in the sum $q+\kappa$ in the denominator of
(\ref{Uscreened-general}), one finds that the exchange interaction
\begin{equation}
{\tilde V}^\rho(i\omega _{m},q)={1\over 2\nu}\
\frac{S-1/\tau}{S-W}
\label{Uscreened-appr}
\end{equation}
has a singularity $\propto 1/q^2$ in the limit $q\to 0$:
\begin{equation}
{\tilde V}^\rho(i\omega _{m},q)=
{1\over 2\nu}\ \frac{2|\omega_{m}|(|\omega_{m}|+1/\tau)}{ q^{2}v_{F}^{2}},
\qquad q\to 0,
\end{equation}
so that each separate term $\Sigma_{ij}^a$,  $\Sigma_{ij}^b$ in the self-energy
would diverge. This divergence is analogous to the one encountered in
course of calculation of the tunneling density of states
\cite{altshuler}. In that case, one has to keep the $q$ term in
denominator of (\ref{Uscreened-general}), which cuts off the
logarithmic divergence. For the present problem, this is, however, not
needed. Indeed, as was emphasized in the end of Sec.~\ref{s2}, the
kernel  $K(i\omega_m, q)$ combining together contributions of all
relevant self-energy diagrams
is proportional to $q^2$ in the limit $q\to 0$
and hence cancels the singularity in ${\tilde V}(i\omega _{m},q)$.
In view of this,  it is convenient to represent the kernel function (\ref{Kwq})
in a form which shows explicitly that $K(i\omega_m, q=0)=0$
[we recall that $S_0(q=0)=|\omega_m|$ and $S(q=0)=W$]:
\be
K(i\omega_m,q)=\left(-{1\over S_0}+{1\over
    |\omega_m|}\right)+\frac{S-W}{S-1/\tau}
\left[\frac{1}{S\tau(S-1/\tau)}-\frac{1}{|\omega_m|}\right].
\ee
Then the product ${\tilde V}^\rho(i\omega _{m},q)K(i\omega_m,q)$ takes the form
\be
2\nu {\tilde V}^\rho(i\omega _{m},q)
K(i\omega_m,q)=\frac{S_0-|\omega_m|}{S_0(S-W)}-\frac{1}{S_0}+\frac{1}{S\tau(S-1/\tau)}.
\label{product-singlet}
\ee
Performing the integration over the momentum $q$ in (\ref{sigmaVK}),
we get the  correction to self-energy in the singlet (charge)
channel
\be
\delta\Sigma^\rho(i\e_n,\xi_0)=-{i\ T\over 2 E_F \tau} \sum_{\omega_m>\e_n}^{\Delta}
\left\{
  (1+2\omega_m\tau)\ln\frac{1+2\omega_m\tau}{2\omega_m\tau}-1+\ln\frac{\Delta^2+\omega_m^2}{\omega_m^2}  \right\},
\label{deltasigma-Coul-sum}
\ee
where we introduced
\be
\Delta\equiv \kappa v_F.
\ee
Comparing (\ref{deltasigma-Coul-sum}) and (\ref{summshort-tot}), we see that
the last term in (\ref{deltasigma-Coul-sum}) corresponds to a static
short-range interaction with $\nu U_0=1$.

Setting $\e_n=\pi T$ and separating the contributions to the sum
(\ref{deltasigma-Coul-sum}) governed by the high-energy
($\e\sim\Delta$) and low-energy ($\e\sim T$) regions, we can present
the result in the following form:
\be
\delta\Sigma^\rho(i\pi T,\xi_0)=-{i\ T\over 2 E_F \tau}\left[{{\rm const} \Delta\over
    T}-\left(1-{1\over 8\pi T\tau}\right) \ln{\Delta\over T}-f(4\pi T\tau) \right],
    \label{coulomb-delta-sigma-result}
\ee
where $f(x)$ is a parameterless function,
\be
\label{fx-crossover}
f(x)=\sum_{m=1}^\infty\left[1 - (1+mx)\ln\frac{1+mx}{mx}+\frac{1}{2mx}
\right]=\left\{
\begin{array}{ll} \displaystyle
{c_{\rm f}\over x}+\left[{1\over 2x}+{1\over 2}\right]\ln{1\over x}, & \quad  x\ll 1,
\\[0.5cm]
\displaystyle
{\pi^2\over 36 \ x^2}, & \quad x\gg 1,
\end{array}
\right.
\ee
with $c_{\rm f}=-3/4 -\psi(1)/2 = -0.461392..$ (here $\psi(x)$ is the digamma function).
Thus the dynamical screening of Coulomb interaction leads
to different asymptotics of the self-energy in the diffusive
and ballistic regimes, in contrast to the case of weak short-range
interaction~\cite{footnote-i}.

\begin{figure}[ht]
 \includegraphics[width=0.45\columnwidth]{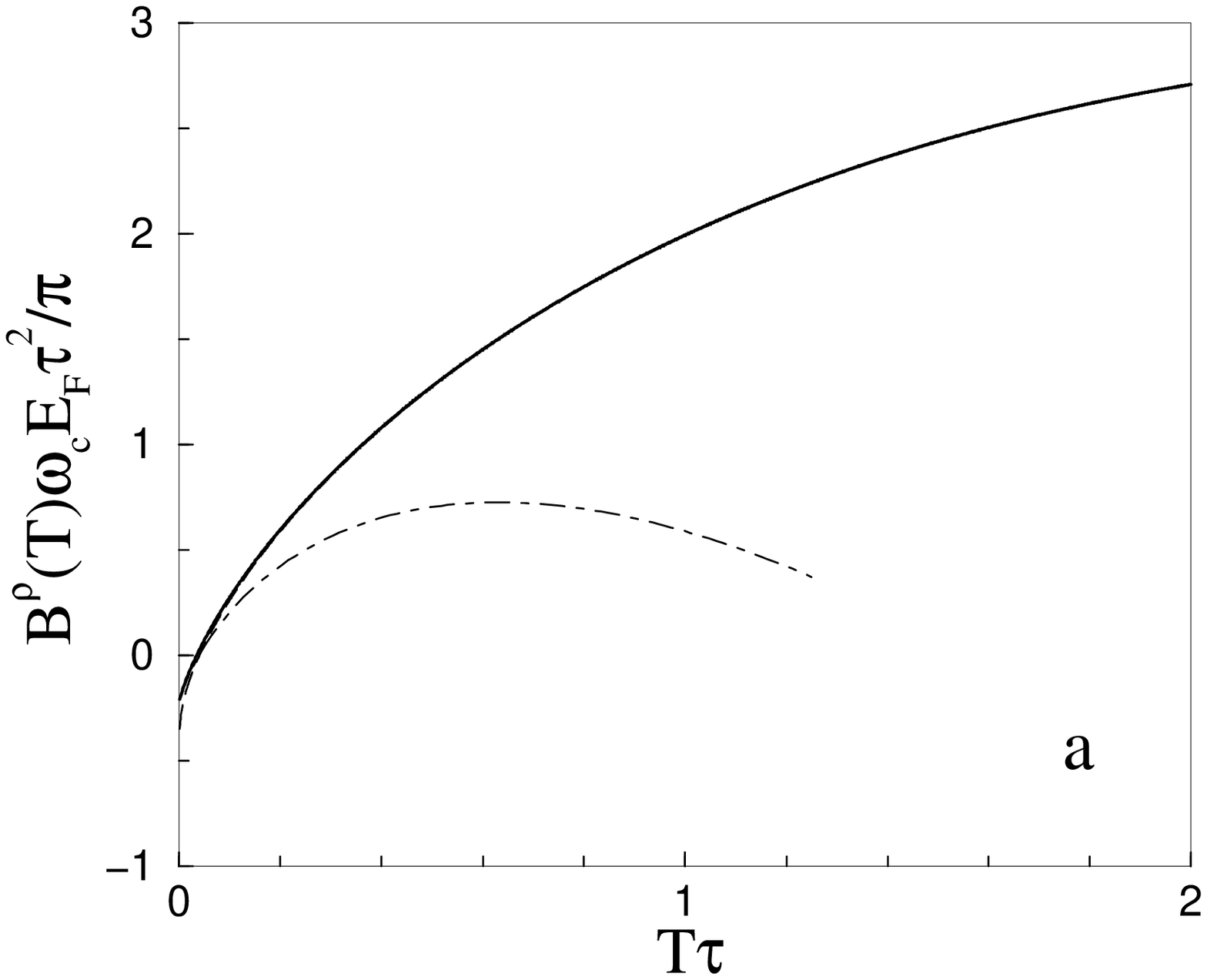} \hskip0.6cm
 \includegraphics[width=0.465\columnwidth]{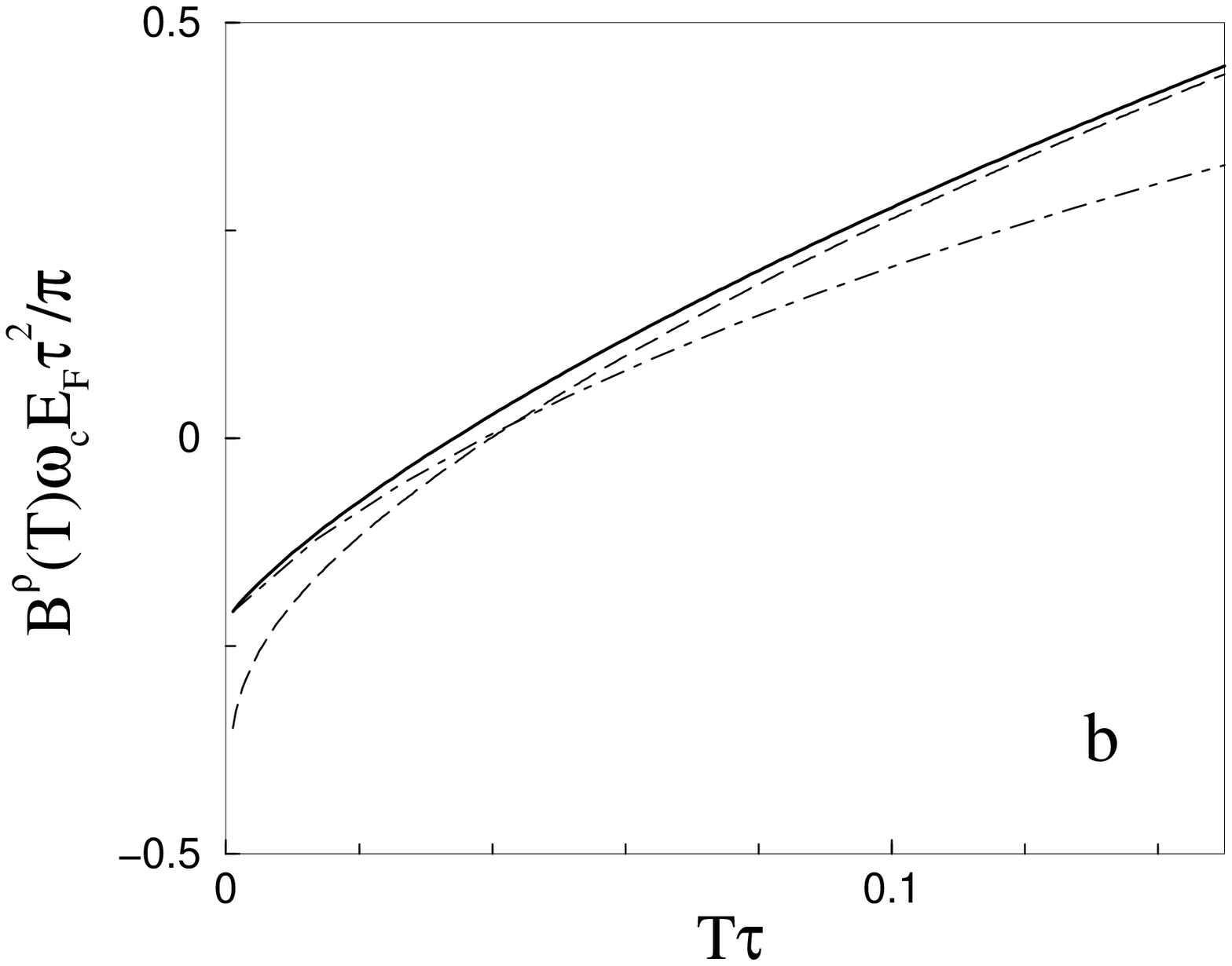}
 \caption{Temperature dependence  of the
 singlet channel correction to the damping factor $B^\rho(T)$,
 Eq.~(\ref{B-rho-T}), for $4\pi \Delta\tau=100$ (solid line)
 with the low-$T$ (dot-dashed) and high-$T$ (dashed) asymptotics,
 Eq.~(\ref{B-rho-T-asympt}). 
 (a) Wide temperature range: on this scale $B^\rho(T)$ is essentially
 indistinguishable from its high-$T$ asymptotics; 
 (b) low-$T$ part: the crossover between the two asymptotics occurs at
 $T\tau\sim 0.05$.} 
 \label{Brho}
\end{figure}

The $T$-dependence of the leading
correction to the magnetooscillations
 damping factor due to the interaction in the singlet channel has
 therefore the form
\bea
\label{B-rho-T}
B^\rho(T)&=&
\frac{\pi}{\omega_{c}\tau}\frac{T}{E_{F}}\left[\left(1-{1\over 8\pi
      T\tau}\right)\ln\frac{\Delta}{T}+ f(4\pi T\tau) \right]
 \\[0.5cm]
&=&\frac{\pi}{\omega_{c}\tau}\frac{T}{E_{F}}  \times  \left\{
\begin{array}{ll} \displaystyle
{3 \over 2}\ln\frac{\Delta}{T} - {1\over 2}\left(1+\frac{1}{4\pi T\tau}\right) \ln(4\pi \Delta\tau)
- \frac{c_{\rm f}}{4\pi T\tau} , & \quad  4\pi T\tau \ll 1,
\\[0.5cm]
\displaystyle
\left(1-\frac{1}{8\pi T\tau}\right) \ln\frac{\Delta}{T}, & \quad 4\pi T\tau \gg 1.
\end{array}
\right.
\label{B-rho-T-asympt}
\eea
This result is illustrated in Fig.~\ref{Brho}. Note that the ballistic
asymptotics describes the exact result with
a remarkable accuracy down to very low temperature, $T\tau\sim
0.01-0.05,$ see Fig.~\ref{Brho}a. 
Retaining in Eq.~(\ref{B-rho-T-asympt}) only the leading terms and
suppressing the $T$-independent 
contributions which can be absorbed in the FL-renormalized $\tau$, we get
\be
B^\rho(T)=\frac{\pi}{\omega_{c}\tau}\frac{T}{E_{F}}  \times  \left\{
\begin{array}{ll} \displaystyle
{3 \over 2}\ln\frac{\Delta}{T} - {1\over 2} \ln(4\pi \Delta\tau) , &
\quad  4\pi T\tau \ll 1, 
\\[0.5cm]
\displaystyle
\ln\frac{\Delta}{T}, & \quad 4\pi T\tau \gg 1.
\end{array}
\right.
\label{B-rho-T-lead-asympt}
\ee

\subsection{Triplet channel}
\label{s5.2}

Calculation of the corresponding triplet contribution $B^\sigma(T)$ 
is presented in
Appendix C and leads to qualitatively similar asymptotics.
The leading term in the total correction to the damping factor in the
ballistic regime, 
realized in experiments on low-disorder
samples at realistic temperatures, takes the simple form
\be
B(T)=B^\rho(T)+B^\sigma(T)\simeq\left(1+\frac{3F_0^\sigma}{1+F_0^\sigma}\right)
\frac{\pi}{\omega_{c}\tau}\frac{T}{E_{F}}\ln\frac{\Delta}{T}.
\label{B-tot-T-lead}
\ee
As discussed in Sec.~\ref{s4},
this result arises due to the correction to the effective mass (the
consideration of Sec.~\ref{s4} fully 
applies to the case of the Coulomb interaction as well).

It is worth noting that due to the factor $(1+F_0^\sigma)$ in the
denominator of $B^\sigma(T)$, the 
damping of magnetooscillations tends to diverge upon approaching the
Stoner ferromagnetic instability.
Since the damping is determined by the effective mass $m^*(T)$ (see
Sec.~\ref{s4}), we conclude that 
$m^*(T)$ diverges as $F_0^\sigma\to -1$ due to the interplay of
disorder and interaction. 
This should be contrasted with the clean case, where the effective
mass is solely determined by 
Fermi-liquid constant in the singlet channel, $F_1^\rho,$
and hence is insensitive to the ferromagnetic instability.

\subsection{Discussion and comparison to earlier work}
\label{s5.3}

Let us discuss the obtained results for the damping factor $B(T)$. 
In both the diffusive and ballistic regimes we find the temperature
dependence of the form $B(T)\propto T\ln T$. In the diffusive (low-$T$)
regime the relative correction to the damping factor,
$\omega_c\tau B \sim g^{-1} T\tau \ln T$ 
is less singular than the known corrections \cite{altshuler}
to the conductivity, $\delta\sigma /\sigma\sim g^{-1}\ln T$, 
and the tunneling density of states,
$\delta\rho/\rho \sim g^{-1}\ln^2 T$ (here $g\sim E_F\tau$ is the
dimensionless conductivity). On the technical level, the qualitative
difference with the tunneling DOS can be traced back to the
contribution of the Hikami box diagrams, $\Sigma_{ij}^{b}$. 

As has been already mentioned in Sec.~\ref{s1}, the problem of the
effect of the interaction on magnetooscilations was recently addressed
in Ref.~\onlinecite{MMR}. The result of this work  
for the damping factor is qualitatively similar to ours, $B\propto
T\ln T$. However, the crossover function and, in particular, the 
prefactors in both ballistic and diffusive limits  differ from ours
(1 instead of 3/2 in the first line of Eq.~(\ref{B-rho-T-lead-asympt}),
and 3/2 instead of 1 in the second line). 
This difference is not surprising, since the
authors of Ref.~\onlinecite{MMR} took into account only one diagram
$\Sigma^a_{01}$ out of six in Fig.~\ref{f-sigma}. Thus, even the
qualitative agreement may be considered as an accidental coincidence. 
In fact, there is a conceptual difference between our result and that
of Ref.~\onlinecite{MMR}. To illustrate this, consider a toy
interaction of the type $V(\omega,{\bf q})= F(\omega)\delta(\bq)$. Our
result then would be zero, since the kernel function $K(i\omega_m,\bq)$,
Eq.~(\ref{Kwq}), satisfies the gauge-invariance constraint 
[see Eq.~(\ref{K(q=0)=0}) and discussion below it],
\be
\label{gauge-con}
K(i\omega_m,\bq)|_{\bq=0}=0.
\ee
In contrast, the formula of  Ref.~\onlinecite{MMR} would give a finite
result, since their kernel 
\be
\label{kernel-MMR}
K_{01}^{(a)} = {2\over S}{1\over S\tau-1}
\ee
does not satisfy the requirement (\ref{gauge-con}),
$$
K_{01}^{(a)}(i\omega_m, q=0) =  {2\over |\omega_m| (|\omega_m|\tau +1)}.
$$
For the Coulomb interaction this results in a logarithmic divergency
at small $q$ that is cut off by the plasmon pole at $q \sim q_{\rm
  min}$ with
$$
q_{\rm  min} = { |\omega_m| (|\omega_m| + 1/\tau) \over \kappa v_F^2}, 
$$
similarly to the calculation of the tunneling density of states
\cite{altshuler}. As we explained in the end of Sec.~\ref{s2}, the
contribution of small momenta, $q< R_c^{-1}$, should be suppressed for
the present problem. Therefore, the above small-$q$ divergence should
be cut off by the magnetic field, which would partly transform 
$\ln T$  of Ref.~\onlinecite{MMR} into $\ln B$. However, this problem
is in fact spurious: the result of our work does not suffer from any
infrared divergencies, since our kernel  $K(i\omega_m,\bq)$ does
satisfy Eq.~(\ref{gauge-con}).

Finally, let us briefly comment on the ultra-diffusive regime, $T\ll \omega_c\ll 1/\tau.$
In this regime the summation over Matsubara energies $\e_n$ is not restricted
to $n=0$ and $N_c\sim \omega_c/T\gg 1$ Matsubara harmonics are important. 
Therefore, the damping due to the inelastic scattering 
[suppressed only for $n=0$, see Ref.~\onlinecite{MMR} and Eq.~(\ref{sigma++FL})]
becomes finite.  The corresponding contribution to the damping can be roughly estimated using
Eq.~(\ref{sigma++FL}) taken at relevant $n\sim N_c$: 
\be
\delta B^{\rm inel}(T)\sim -\left.n^2{T^2\over \omega_c E_F}\ln(E_F\tau)\right|_{n\sim N_c} 
\sim -{\omega_c\over E_F}\ln(E_F\tau),
\ee
yielding $|\delta B^{\rm inel}(T)|\ll 1$, since in the ultra-diffusive regime 
$\omega_c\ll 1/\tau$.
Thus the inelastic contribution to the damping factor is always small.
Note that the contribution to the damping due to the renormalization of the 
effective mass in the ultra-diffusive regime is also small:
at $n=N_c$ we have $B(T)=-2\pi \e_n/\omega_c \delta m/m \sim \delta m/m \ll 1.$

\section{Conclusions}
\label{s6}

In conclusion, we have studied the $T$-dependent interaction corrections
to the damping of magnetooscillations in a two-dimensional electron
gas.  The damping factor has been calculated for Coulomb
and short-range interaction in the whole range of temperatures,
from the ballistic to the diffusive regime. While the relevant
diagrams are similar to those for the local density of states, the
results are essentially different, see Eqs.~(\ref{FT-short-result}),
(\ref{B-rho-T}).

We have identified leading contributions to the
damping induced by interplay of interaction and disorder, which
can be associated with corrections to the effective mass and
the quantum scattering time.   It has been shown that in the
ballistic regime, which is typically realized in low-disorder
samples at realistic temperatures, the dominant
effect is that of the renormalization of the effective electron mass
due to the interplay of the interaction and impurity scattering.
Specifically, the correction to the effective mass is of the form
$\delta m/m\sim 1/(E_F\tau) \ln (E_F/T)$, Eq.(\ref{deltamassT-result}). The
correction to the impurity scattering time is of the form
$\delta\tau_q/\tau \sim T/E_F$, Eq.~(\ref{deltatauT-result}),
and yields a subleading contribution to the damping.  We have
confirmed the result for the correction to the quantum scattering time
by performing a calculation based on the picture of scattering by
impurities dressed by Friedel oscillations. The
results of the paper are relevant to the analysis of experiments on
magnetooscillations (in particular, for extracting the value of the
effective mass) and are expected to be useful for understanding the
physics of a high-mobility 2DEG near the apparent metal-insulator
transition.

\section{Acknowledgments}

We thank I.~Aleiner, D.~Aristov, A.~Chubukov, S.~Kravchenko, and V.~Pudalov
for useful discussions.
We are particularly grateful to D.~Maslov and M.~Reizer for discussions
of Ref.~\onlinecite{MMR} and comments on the draft of our paper.
This work was supported by SPP ``Quanten-Hall-Systeme'' and CFN of DFG,
by US DOE under contract No. DE-AC02-98CH1-886, by the Program ``Russian Science
School'' under grant No. 2192.2003.2, and by RFBR.

\appendix

\section{Luttinger formalism for the thermodynamic potential}
\label{App1}
\renewcommand{\theequation}{A.\arabic{equation}}
\setcounter{equation}{0}

In this Appendix we derive the oscillatory part of the thermodynamic potential $\Omega$,
following Ref.~\onlinecite{luttinger}.

We calculate the sum in Eq.~(\ref{Omega-trace}), using Poisson's summation
formula
\be
\sum _{m=0}^{\infty }f(m\omega _{c})=\frac{1}{2}f(0)+
\int _{0}^{\infty } \frac{d\xi }{\omega _{c}} \ f(\xi ) -
\sum ^{\infty }_{k=1}\frac{1}{2\pi ik} \int _{0}^{\infty }d\xi \
f'(\xi ) \ \left (e^{2\pi ik \xi/\omega_{c}}-
e^{-2\pi ik \xi/\omega_{c}}\right).
\label{sumPoisson}
\ee
From Eq.~(\ref{Omega-trace}) we have $f'(\xi)=-G(i\e_n,\xi)$, where $G(i\e_n,\xi)$ is the Green's function.
Thus, to extract the oscillatory contribution to $\Omega$, we should calculate the following integral
\be
I_{\pm}=\int _{0}^{\infty }d\xi \
\frac{e^{\pm 2\pi ik \xi/\omega_{c}}}{\xi -\mu
  -i\varepsilon_{n}+\Sigma_{\rm ee}(i\e_n,\xi)-
i\sgn\e_n/2\tau}
\left[1+ \frac{\partial\Sigma_{\rm ee}(i\e_n,\xi)}{\partial \xi }\right] .
\label{Integral}
\ee
Here we introduce the self-energy $\Sigma_{\rm ee}(i\e_n,\xi)$,
which is a function defined in the plane of a complex variable
$\xi$, such that
\begin{equation}
\Sigma_{\rm ee}(i\e_n,m\omega_c)=\Sigma_{\rm ee}(i\e_n,\xi=m\omega_c).
\label{self-energy-m-to-xi}
\end{equation}

The main contribution comes from the pole $\xi=\xi_0$, where
$\xi_0$ obeys the self-consistent equation
\begin{equation}
\xi_0=\mu+i\e_n+{i\over 2\tau}\sgn\e_n-\Sigma_{\rm ee}(i\e_n,\xi_0).
\label{self-consistent-xi0}
\end{equation}

We expand the self-energy (from now on we will skip the subscript
``${\rm ee}$" in $\Sigma_{\rm ee}$)
in the vicinity of the pole,
\begin{equation}
\Sigma(i\e_n,\xi)\simeq \Sigma(i\e_n,\xi_0)+(\xi-\xi_0)
\left.\frac{\partial \Sigma (i\e_n,\xi )}{\partial \xi }\right|_{\xi=\xi_0}.
\label{self-energy-expanded}
\end{equation}
Then the denominator in Eq.~(\ref{Integral}) becomes proportional
to $(1+\partial\Sigma/\partial\xi)$ so that
these factors drop out and the integral takes a simple form
\begin{equation}
I_{\pm}=\int _{0}^{\infty }d\xi \frac{\exp[\pm 2\pi ik\ \xi/\omega_c]}{\xi -\xi_0}.
\label{Integral-simple}
\end{equation}

We first single out the FL renormalization factors in $\Sigma$, i.e.
represent the self-energy in the following form
(assuming a constant electron concentration)
\begin{equation}
\Sigma(i\varepsilon_n,\xi) \simeq \delta\mu+
\beta_0(\xi-\tilde\mu)-i\alpha_0\e_n+\delta\Sigma(i\varepsilon_{n},\xi ),
\label{Sigma00-full}
\end{equation}
where
$$\tilde\mu=\mu-\delta\mu=\pi n_e/m$$
is the chemical potential for noninteracting electrons
($n_e$ is the electron concentration):
$\xi-\tilde\mu \simeq v_F(k-k_F).$ The correction to the self-energy
$\delta\Sigma(i\varepsilon_{n},\xi)$ contains contributions that are
smaller that $\beta_0(\xi-\mu)$ and $\alpha_0\e_n$ by either $T/E_F$
or $1/E_F\tau$. These additional contributions are related to the inelastic processes
and to the modification of the pure FL result due to disorder.

In order to solve Eq.~(\ref{self-consistent-xi0}), we treat the subleading terms constituting
$\delta\Sigma$ as small corrections.
Solving Eq.~(\ref{self-consistent-xi0}) by iterations, we first find
its solution neglecting these small corrections
and expanding $\Sigma_{00}$ around the mass-shell:
\begin{equation}
\xi_0^{(0)}\simeq\tilde\mu+i\e_n+{i\over 2\tau}\sgn\e_n +
i\alpha_0\e_n - \beta_0(\xi_0^{(0)}-\tilde\mu),
\label{self-cons-xi00}
\end{equation}
which yields
\begin{equation}
\xi_0^{(0)}= {\pi n_e\over m} + i\e_n\frac{1+\alpha_0}{1+\beta_0}+
{i\sgn\e_n\over 2\tau(1+\beta_0)}.
\label{xi00}
\end{equation}
Next we use this value of $\xi_0$ in $\delta\Sigma$
and solve the self-consistent equation again, now keeping the terms
previously neglected. Then we arrive at
\begin{equation}
\xi_0\simeq \xi_0^{(0)}-\delta\Sigma\left(i\e_n,\xi_0^{(0)}\right)/(1+\beta_0).
\label{xi0result}
\end{equation}

Substituting this value of the pole in Eq.~(\ref{Integral-simple}), we obtain
\begin{equation}
\Omega =\Omega _{\rm osc}+\Omega _{\rm Non-osc},
\end{equation}
where
\begin{eqnarray}
   \Omega_{\rm Non-osc} &=&-T\, \nu\, \omega_{c}\sum^{\infty }_{n=-\infty}
   \ln\left(
    -\mu -i\varepsilon_{n}+
    \Sigma(i\varepsilon_{n},0)-\frac{i\sgn\e_n}{2\tau}
  \right)\nonumber\\
 &-&2\nu T\sum_{n=-\infty}^{\infty }\int_{0}^{\infty } d\xi
 \ln \left(
    \xi -\mu
    -i\varepsilon_{n}+\Sigma(i\e_n,\xi)
    -\frac{i\sgn\e_n}{2\tau}
  \right)
  \label{eq-omega-Non-osc}
\end{eqnarray}
and
\begin{eqnarray}
    \Omega_{\rm osc}&=&\sum _{k=1}^{\infty }\sum ^{\infty }_{n=-\infty}
  \frac{T\, 2\nu\, \omega_{c}}{k}
\exp \left\{
    \frac{2\pi k}{\omega_c}
    \left(
      i {\pi n_e\over m} \sgn\e_n
       - {|\e_n|(1+\alpha_0)+1/2\tau+
i\delta\Sigma(i\e_n,\xi_0)\sgn\e_n
\over 1+\beta_0}
\right)
  \right\}\nonumber \\
&=&
\sum _{k=1}^{\infty}\, {4\nu\  T \omega_c\over k}\, \cos{2\pi^2 k n_e \over e B}
\sum ^{\infty }_{n=0}
\exp \left\{
    -\frac{2\pi k }{\omega_c}\frac{1+\alpha_0}{1+\beta_0}
    \left[
    \e_n+{1\over 2\tau (1+\alpha_0)} +
      \frac{i\delta\Sigma (i\e_n,\xi_0)}{1+\alpha_0}\right]
      \right\}\\
&\simeq&
2\nu\left({\omega_c\over 2\pi} \right)^2
A_1\cos{2\pi^2  n_e \over e B}.
\label{eq-omega-osc-k}
\end{eqnarray}
Here $A_1$ is the amplitude of the principal harmonics of the
oscillations,
\bea
A_1&\equiv&\frac{4 \pi^2 T}{\omega_c}\sum_{\e_n>0}
  \exp\left(
    -\frac{2\pi }{\omega_c}
    \frac{1+\alpha_0}{1+\beta_0}
    \left[
    \e_n+{1\over 2\tau(1+\alpha_0)} +
      {i\delta\Sigma (i\e_n,\xi_0)\over 1+\alpha_0} \right]
  \right).
\end{eqnarray}

\section{Calculation of self-energies}
\label{AAGF}
\renewcommand{\theequation}{B.\arabic{equation}}
\setcounter{equation}{0}

In this Appendix we calculate the relevant self-energy contributions and
derive Eq.~(\ref{Kwq}).
The zero-$B$ Green's function is given by
\begin{eqnarray}
G(i\varepsilon_n-i\omega_m,\vec{p}-\vec{q})&=&\frac{i}{(\omega_{m}-\varepsilon_{n})(1+\alpha_0)+
  \sgn(\omega_m-\varepsilon_n)/2\tau -[i(\xi-{\tilde \mu})
    -iv_{F}q\cos\theta](1+\beta_0)}\nonumber \\
&=&\frac{i}{{\tilde\omega_{m}}-{\tilde\varepsilon_{n}}+
  \sgn(\omega_m-\varepsilon_n)/2\tau -i(\xi-{\tilde \mu})(1+\beta_0)
    +iv_{F}{\tilde q}\cos\theta}.
    \label{FLGF}
\end{eqnarray}
We denote the FL-renormalized energies and momenta as
${\tilde\omega_{m}}=(1+\alpha_0)\omega_{m},\
{\tilde\varepsilon_{n}}=(1+\alpha_0)\varepsilon_{n},$
and ${\tilde q}=(1+\beta_0)q$.
As the first approximation, we have
set in Eq.~(\ref{FLGF}) $p/m=v_{F}$ in the linear-in-$q$ term $iv_{F}q\cos\theta$
and neglected $q^2/2m$.

Since the effective interaction (\ref{Vrho}) and (\ref{Vsigma}) does not depend
on the polar angle of the transferred momentum $\bq$, we average
the FL-dressed Green's function
$G(i\varepsilon_n-i\omega_m,\vec{p}-\vec{q})$
over the angle between $\bp$ and $\bq$.
The result of angle-averaging is
\begin{eqnarray}
  \label{eq:intG}
  \langle G(i\varepsilon_n-i\omega_m,\vec{p}-\vec{q})\rangle&=&
  \int {d\theta\over 2\pi} G(i\varepsilon_n-i\omega_m,\vec{p}-\vec{q})\\
    &=&\frac{i \, \sgn(\omega_m-\varepsilon_n)}{
    \sqrt{
      [\tilde\omega_{m}-\tilde\varepsilon_{n}+\sgn(\omega_m-\varepsilon_n)/2\tau
      -i(\xi-{\tilde \mu})(1+\beta_0)]^{2}+v_{F}^{2}{\tilde q}^{2}}
    }.
\label{fGaveroverphi}
\end{eqnarray}

We substitute $\xi=\xi_0\simeq \xi_0^{(0)}={\tilde \mu}+i\tilde\e_n/(1+\beta_0)+i\sgn \e_n/2\tau(1+\beta_0)$
for $\xi$ in (\ref{fGaveroverphi}) since we are interested in
$\delta\Sigma(i\e_n,\xi_0)$ (the only place where $\xi_0$
appears in $\delta\Sigma$ is the Green's function under the interaction line).
Then the denominator in (\ref{fGaveroverphi}) for $q=0$ reads
\begin{eqnarray}
&&\tilde\omega_{m}-\tilde\varepsilon_{n}+
\sgn(\omega_m-\varepsilon_n)/2\tau -i(\xi-{\tilde \mu})(1+\beta_0)\nonumber\\
&&=\tilde\omega_{m}-\tilde\varepsilon_{n}+
\sgn(\omega_m-\varepsilon_n)/2\tau-i(1+\beta_0)
\left[i{\tilde\e_n\over 1+\beta_0}+i{\sgn \e_n\over 2\tau(1+\beta_0)}\right]
\nonumber\\
&&=\tilde\omega_{m}+{1\over \tau}\theta(\omega_m-\varepsilon_n)\theta(\varepsilon_n),
\end{eqnarray}
with $\theta(x)$ the theta-function.
For definiteness, below we consider $\e_n>0$.

For $\Sigma_{01}$ and $\Sigma_{11}$ at $\e_n>0$
we consider the Green's function at $\omega_m-\e_n>0$
in order to have different signs of Matsubara energies in the
Green's functions connected by the interaction vertex.
This condition allows us to dress the interaction vertices by impurity
ladders.

We see that $\e_n$ drops out in the averaged Green's function
taken at $\xi=\xi_0,$ as in two-particle quantities:
\begin{equation}
\langle G(i\varepsilon_n-i\omega_m,\vec{p}-\vec{q})\rangle\left.\right|_{\xi=\xi_0}=
\frac{i}{
    \sqrt{
      (|\tilde\omega_m|+1/\tau)^{2}+v_{F}^{2}{\tilde q}^{2}}
    }, \qquad \omega_m>\e_n.
\label{ball-diff_RA}
\end{equation}
When both $\e_n$ and $\e_n-\omega_m$ have the same sign
(such a contribution appears in the calculation
of $\delta\Sigma_{00}$), we find
\begin{equation}
\langle G(i\varepsilon_n-i\omega_m,\vec{p}-\vec{q})\rangle\left.\right|_{\xi=\xi_0}
= -\frac{i}{
    \sqrt{
      |\tilde\omega_m|^{2}+v_{F}^{2}{\tilde q}^{2}}
    }, \qquad \omega_m<\e_n.
\label{ball-diff_RR}
\end{equation}

Now we re-define the Fermi velocity to absorb the
FL-factors according to
\begin{equation}
v_F^*=v_F\frac{1+\beta_0}{1+\alpha_0}={k_F\over m^*}.
\end{equation}
Then we can return from $\tilde\omega_m$ and $\tilde q$ to
$\omega_m$ and $q$, expressing the angle-averaged
Green's function in terms of FL-renormalized parameters
$Z$, $\tau^*$ [introduced in Eq.~(\ref{taustar})]
and $v_F^*$:
\begin{eqnarray}
\langle G(i\varepsilon_n-i\omega_m,\vec{p}-\vec{q})\rangle\left.\right|_{\xi=\xi_0}
&=& \frac{i Z}{
    \sqrt{
      (|\omega_m|+1/\tau^*)^{2}+(v_{F}^*q)^{2}}
    }, \qquad \omega_m>\e_n, \nonumber \\
\langle G(i\varepsilon_n-i\omega_m,\vec{p}-\vec{q})\rangle\left.\right|_{\xi=\xi_0}
&=& -\frac{i\ Z }{
    \sqrt{
      |\omega_m|^{2}+(v_{F}^* q)^{2}}
    }, \qquad \omega_m<\e_n.
    \nonumber
\end{eqnarray}
Furthermore, the $Z$-factor will be cancelled in the final result,
when $\langle G(i\varepsilon_n-i\omega_m,\vec{p}-\vec{q})\rangle$
is used to calculate the correction to the observables, see e.g. Ref.~\onlinecite{Falko}
and discussion in Sec.~IIB.

Using (\ref{ball-diff_RA}) and (\ref{ball-diff_RR}), we obtain
\bea
\Sigma_{01}^{a}(i\e_n,\xi_0)&=&
-T\sum_{\omega_{m}>\varepsilon_{n}}\int
\frac{d^{2}q}{\left( 2\pi \right) ^{2}}\,V(i\omega_m,\bq)\Gamma(i\omega_m,\bq)
\left. \langle G(i\e_n,\bp-\bq)\rangle\right|_{\xi_{p}=\xi_{0}}
\nonumber \\
&=&-i\ T\sum_{\omega_{m}>\varepsilon_{n}}\int
\frac{d^{2}q}{\left( 2\pi \right)^{2}}\,
\,\frac{V(i\omega_m,\bq)\, \Gamma(i\omega_m,\bq)}{S(i\omega_m,\bq)},
\label{deltasigma01a}
\\
\Sigma_{11}^{a}(i\e_n,\xi_0)&=&
-T\sum_{\omega_{m}>\varepsilon_{n}}\int
\frac{d^{2}q}{\left( 2\pi \right) ^{2}}\,V(i\omega_m,\bq)\Gamma^2(i\omega_m,\bq)
\left. \langle G(i\e_n,\bp-\bq)\rangle\right|_{\xi_{p}=\xi_{0}}
\nonumber \\
&=&-i\ T\sum_{\omega_{m}>\varepsilon_{n}}\int
\frac{d^{2}q}{\left( 2\pi \right)^{2}}\,
\,\frac{V(i\omega_m,\bq)\,\Gamma^2(i\omega_m,\bq)}{S(i\omega_m,\bq)},
\label{deltasigma11a}
\eea
where for brevity we introduce new variables
$
S(i\omega_m,\bq)\equiv\sqrt{(|\omega_m|+1/\tau)^{2}
+v_{F}^{2}q^{2}}=\sqrt{W^{2}+v_{F}^{2}q^{2}}
$
and
$
W\equiv|\omega_m|+1/\tau.
$

Now we consider the contribution to the self-energy
without vertex corrections, $\Sigma_{00}^a(i\e_n,\xi)$.
We recall that in $\Sigma_{00}^a(i\e_n,\xi)$ the summation over
transferred frequencies is not restricted to $\omega_m>\e_n$.
Presenting  $\Sigma_{00}^a(i\e_n,\xi)$ as
\bea
\Sigma_{00}^a(i\e_n,\xi_0)&=&\Sigma_{00}^{a,+-}(i\e_n,\xi_0)+
\Sigma_{00}^{a,++}(i\e_n,\xi_0),
\\
\Sigma_{00}^{a,+-}(i\e_n,\xi_0)&=&
-T\sum_{\omega_{m}>\varepsilon_{n}}\int
\frac{d^{2}q}{\left( 2\pi \right) ^{2}}\,V(i\omega_m,\bq)
\left. \langle G(i\e_n,\bp-\bq)\rangle\right|_{\xi_{p}=\xi_{0}}
\nonumber \\
&=&-i\ T\sum_{\omega_{m}>\varepsilon_{n}}\int
\frac{d^{2}q}{\left( 2\pi \right)^{2}}\,
\frac{V(i\omega_m,\bq)}{S(i\omega_m,\bq)},
\eea
we further split the contribution $\Sigma_{00}^{a,++}(i\e_n,\xi_0)$
corresponding to no change of Matsubara frequencies at the interaction
vertices
into two parts as follows
\bea
\Sigma_{00}^{a,++}(i\e_n,\xi_0)&=&
-T\sum_{\omega_{m}<\varepsilon_{n}}\int
\frac{d^{2}q}{\left( 2\pi \right) ^{2}}\,V(i\omega_m,\bq)
\left. \langle G(i\e_n,\bp-\bq)\rangle\right|_{\xi_{p}=\xi_{0}}
\nonumber \\
&=&
i\ T \int
\frac{d^{2}q}{\left( 2\pi \right)^{2}}\,
\left\{
\sum_{|\omega_{m}|<\varepsilon_{n}}
+ \sum_{\omega_{m}<-\varepsilon_{n}}\right\}
\frac{V(i\omega_m,\bq)}{\sqrt{|\omega_m|^2+q^2v_F^2}}
\label{sigma00}
\eea
The terms with $|\omega_{m}|<\varepsilon_{n}$ are responsible for the
FL-renormalization and for
the inelastic (determined by real processes)
contribution to the self-energy, yielding the following FL-type
term~\cite{MMR}:
\be
\Sigma^{\rm FL}(i\e_n,\xi_0)= - i\alpha_0\ \e_n
- i\gamma(i\e_n,T) \frac{\e_n^2-\pi^2T^2}{E_F},
\label{sigma++FL}
\ee
where the function $\gamma(i\e_n,T)$ depends logarithmically on ${\rm max}[\epsilon_n,T,1/\tau]$: 
in particular, $\gamma(i\e_n,T)\propto\ln[E_F/(-i\epsilon_n)]$ for $\epsilon_n\gg T,1/\tau$
and $\gamma(i\e_n,T)\propto\ln[i E_F\tau]$ for $1/\tau\gg T,\epsilon_n$.
The first term in (\ref{sigma++FL}) determines the FL $Z$-factor and has been
separated from $\delta\Sigma$ which governs the correction to the damping factor,
see Eq.~(\ref{Sigma00-full}).
As for the second term, its imaginary part describes the inelastic 
electron-electron scattering,
while its real part contributes to the renormalization
of the effective mass~\cite{chubukov1,chubukov,DasSarma}. However,
when taken at $\epsilon_0=\pi T,$ as appropriate for the damping of the
magnetooscillations at $T\gg \omega_c$, the second term in (\ref{sigma++FL})  vanishes,
in agreement with Ref.~\onlinecite{MMR}.
Note that for the case of weak short-range interaction,
$V(i\omega_m,\bq)={\rm const}(\omega_m),$
the inelastic contribution is zero to the first order in $V$.

Thus in order to calculate $\delta\Sigma,$ we shall retain in (\ref{sigma00})
only the term corresponding to the summation
over $\omega_{m}<-\varepsilon_{n}$.
In this term we
change the sign of $\omega_m$ and (suppressing the irrelevant inelastic term)
and obtain
\be
\Sigma_{00}^{a,++}(i\e_n,\xi_0)=
i\ T\sum_{\omega_{m}>\varepsilon_{n}}\int
\frac{d^{2}q}{\left( 2\pi \right)^{2}}\,
\frac{V(i\omega_m,\bq)}{\sqrt{|\omega_m|^2+q^2v_F^2}},
\ee
thus arriving at
\be
\Sigma_{00}^{a}(i\e_n,\xi_0)=
-i\ T\sum_{\omega_{m}>\varepsilon_{n}}\int
\frac{d^{2}q}{\left( 2\pi \right)^{2}}\,
V(i\omega_m,\bq)\left[\frac{1}{S(i\omega_m,\bq)}- \frac{1}{S_0(i\omega_m,\bq)}\right].
\label{deltasigma00a}
\ee
Here we use the fact that $V(i\omega_m,\bq)=V(-i\omega_m,\bq)$
as the dynamically screened interactions depends on $\omega_m$
only through $|\omega_m|$ and introduced $S_0(i\omega_m,\bq)=\sqrt{\omega_m^2+q^2v_F^2}$.

Let us turn now to the Hikami-box diagrams, shown in Fig.~\ref{f-sigma}b
We remind the reader that these diagrams are generated by covering
each contribution to the self-energy $\Sigma_{ij}$ from
Fig.~\ref{f-sigma}a
with $\e_n(\e_n-\omega_m)<0$
by a single impurity line. Therefore for white-noise disorder (addressed in this paper)
the Hikami-box contribution to the self-energy is independent
of $\xi_p$ and can be expressed through the corresponding $\Sigma_{ij}^a$ as
\bea
\Sigma^b_{ij}(i\epsilon_{n})&=&\frac{1}{2\pi\nu\tau}\int d\xi_{p}\nu
[G(i\epsilon_{n},\xi_{p})]^{2}
\Sigma_{ij}^a(i\epsilon_{n},\xi_{p})
\label{Hikami1}
\\
&=&\frac{i}{\tau}\frac{\partial}{\partial\xi_{p}}
\left.\Sigma_{ij}^a(i\epsilon_{n},\xi_{p})\right|_{\xi_{p}=\xi_{0}}
\eea
\be
\Sigma_{ij}^{b}(i\epsilon_{n})=\frac{i}{\tau}\frac{\partial}{\partial\xi_{p}}
\left.\Sigma_{ij}^a(i\epsilon_{n},\xi_{p})\right|_{\xi_{p}=\xi_{0}}
\label{Hikami2}
\ee
since the pole of $G$ is given by $\xi_{0}$, see Eq.~(\ref{xi0result}).
Differentiating (\ref{fGaveroverphi}) we get
\be
\left.\left\langle\frac{\partial}{\partial\xi_{p}}G(i\varepsilon_{n}-i\omega_{n},\bp-\bq)\right\rangle\right|_{\xi_{p}=\xi_{0}}
=-\frac{|\omega_{m}|+1/\tau}{\{(|\omega_{m}|+1/\tau)^{2}+v_{F}^{2}q^{2}\}^{3/2}}=
-\frac{W}{S^3}.
\ee
This yields
\bea
\Sigma_{00}^{b}(i\e_n,\xi_0)&=&
i\ T\sum_{\omega_{m}>\varepsilon_{n}}\int
\frac{d^{2}q}{\left( 2\pi \right)^{2}}\,
\,\frac{V(i\omega_m,\bq)\,  W}{\tau S^3(i\omega_m,\bq)}
\label{deltasigma00b}
\\
\Sigma_{01}^{b}(i\e_n,\xi_0)&=&
i\ T\sum_{\omega_{m}>\varepsilon_{n}}\int
\frac{d^{2}q}{\left( 2\pi \right)^{2}}\,
\,\frac{V(i\omega_m,\bq)\, \Gamma(i\omega_m,\bq)\, W}{\tau S^3(i\omega_m,\bq)}
\label{deltasigma01b}
\\
\Sigma_{11}^{b}(i\e_n,\xi_0)&=&
i\ T\sum_{\omega_{m}>\varepsilon_{n}}\int
\frac{d^{2}q}{\left( 2\pi \right)^{2}}\,
\,\frac{V(i\omega_m,\bq)\,\Gamma^2(i\omega_m,\bq)\, W}{\tau S^3(i\omega_m,\bq)}
\label{deltasigma11b}
\eea

Combining all the contributions together, we arrive at Eq.~(\ref{Kwq}).

\section{Coulomb interaction: Triplet channel}
\label{a3}
\renewcommand{\theequation}{C.\arabic{equation}}
\setcounter{equation}{0}

In this Appendix we calculate the contribution of the triplet channel
to the damping of magnetooscillations.
The effective interaction in the triplet channel can be found by
replacing $V_0(q)\to F_0^\sigma/2\nu$
in the expression for the singlet channel:
\begin{eqnarray}
2\nu \, V^{\sigma}(i\omega _{m},q)&=&\frac{F_0^\sigma}{1+F_0^\sigma\Pi (i\omega _{m},q)}
=\frac{F_0^\sigma}{1+F_0^\sigma[1-|\omega_m|/(S-1/\tau)]}\nonumber \\
&=&
\frac{F_0^\sigma(S-1/\tau)}{(1+F_0^\sigma)(S-1/\tau)-|\omega_m| F_0^\sigma}
= \frac{F_0^\sigma}{1+F_0^\sigma}\frac{S-1/\tau}{S-w},
\label{Uscreened-general-triplet}
\end{eqnarray}
where we introduced
\be
w\equiv |\omega_m| \frac{F_0^\sigma}{1+F_0^\sigma}+{1\over \tau}.
\ee
This yields
\be
2\nu \, V^{\sigma}(i\omega _{m},q) K(i\omega _{m},q)=\frac{F_0^\sigma}{1+F_0^\sigma}
\left[\frac{S_0-(S-1/\tau)}{S_0(S-w)}
+\frac{S-W}{S-w}\frac{1}{S\tau(S-1/\tau)}\right].
\label{product-triplet}
\ee
Performing the integration over $q$ in (\ref{sigmaVK}) and taking into
account the three triplet terms 
corresponding to different projection of the total spin on the $z$-axis $S_z=0,\pm 1$,
we obtain
\bea
\delta\Sigma^\sigma(i\pi T,\xi_0)&=&-{i\ T\over 2 E_F
  \tau}\frac{3F_0^\sigma}{1+F_0^\sigma} 
\sum_{\omega_m=2\pi T}^{\Delta}
\left\{\left(\frac{\omega_m}{\sqrt{\Delta^2+\omega_m^2}}-1\right)
+\ln\frac{\Delta^2+\omega_m^2}{\omega_m^2}+h(\omega_m\tau,F_0^\sigma)
\right\}, 
\label{deltasigma-Coul-triplet-sum}\\
h(z,y)&\simeq&\left(1-{1\over y}\right)\,
\ln\left[1+y\right]
+
\frac{\ y}{1+y}\ z\ h_1(x,y)
\label{hwF}
\\[0.5cm]
h_1(z,y)&=&\ln\frac{(1+2z)(1+y)}{2z}+\frac{z y+(1+y)}{\sqrt{|2(1+y)z-(z y)^2|}}
\nonumber \\[0.5cm]
&\times& \left\{\,
\begin{array}{ll} \displaystyle
\left[\arcsin\frac{z y-1}{\sqrt{2z+1}}
-\arcsin\frac{z y+(1+y)}{(1+y)\sqrt{2z+1}}
\right],
& \displaystyle\quad  z < \frac{2(1+y)}{y^2},
\\[0.5cm]
\displaystyle
\left[
\ln\left(1+\frac{2+y}
{z y-1+\sqrt{(z y)^2-2(1+y)z}}
\right)
-\ln(1+y)
\right]
, & \displaystyle\quad z > \frac{2(1+y)}{y^2}.
\end{array}
\right.
\nonumber
\\
\label{h1xF}
\eea
We see that the first two terms in (\ref{deltasigma-Coul-triplet-sum}) correspond to the
point-like interaction with $\nu U_0 \to 3F_0^\sigma/(1+F_0^\sigma)$,
see Eq.~(\ref{summshort-tot}).
The term $h(\omega_m,F_0^\sigma)$ corresponds to the crossover
function $f(x)$ in the singlet channel,
see Eqs.~(\ref{coulomb-delta-sigma-result}) and (\ref{fx-crossover}).
The result for the singlet channel is reproduced in the limit
$F_0^\sigma\to \infty$ (cf. Ref.~\onlinecite{ZNA}).

The summation over Matsubara frequencies leads to
\be
\delta\Sigma^\sigma(i\pi T,\xi_0)=-{i\ T\over 2 E_F \tau}\frac{3F_0^\sigma}{1+F_0^\sigma}
\left\{ {{\rm const} \Delta\over T}
-\left[1-{\lambda(F_0^\sigma)\over 8\pi T\tau}\right]\ \ln{\Delta\over
  T}-f_\sigma(4\pi T\tau,F_0^\sigma)  \right\},
\label{sum-triplet}
\ee
where
\be
\lambda(y)=\frac{1+y}{y^3}\left[ y(6+y)-2(3+2y)\ln(1+y)\right]
\ee
and
\be
f_\sigma(x,y)=\sum_{m=1}^\infty \left\{ - h\left({m x \over
      2},y\right)+\left[1-\frac{2\ln(1+y)}{y}\right] +
  \frac{\lambda(y)}{2 m x} \right\}
\ee

As a result we obtain the following $T$ dependence of the
triplet contribution to the damping exponent
\be
B^\sigma(T)= \frac{\pi}{\omega_{c}\tau}\frac{3F_0^\sigma}{1+F_0^\sigma}
\frac{T}{E_{F}}\left\{\left[1-{ \lambda(F_0^\sigma)\over 8\pi
      T\tau}\right]\ln\frac{\Delta}{T}+ f_\sigma(4\pi
  T\tau,F_0^\sigma) \right\}.
  \label{B-sigma-result}
  \ee
For not too strong interaction, $(F_0^\sigma)^2/(1+F_0^\sigma)\alt 1,$
(i.e. for $|F_0^\sigma|\alt 0.6,$ 
which is typically met
in experiments, see e.g. Ref.~\onlinecite{sav}) the crossover function
$f_\sigma(x,y)$ 
only yields the subleading $T$-dependence~\cite{footstoner}
of $B^\sigma(T)$, so that the leading contribution to the damping
is given by Eq.~(\ref{FT-short-result}) for the short-range
interaction with $\nu U_0$ replaced by 
$3 F_0^\sigma/(1+F_0^\sigma)$.


\begin{thebibliography}{10}

\bibitem[$^\dagger$]{byline} Also at A.F.~Ioffe
Physical-Technical Institute, 194021 St.~Petersburg, Russia.

\bibitem[$^\#$]{byline1} Also at Petersburg Nuclear Physics
Institute, 188350 St.~Petersburg, Russia.


\bibitem{altshuler} B.L.~Altshuler and A.G.~Aronov, in {\em
    Electron-electron interactions
in disordered conductors}, edited by A.L.~Efros and M.~Pollak
    (Elsevier, 1985), p. 1.

\bibitem{finkelstein} for a review see A.M.~Finkelstein,
  Sov. Sci. Rev. A. Phys {\bf 14}, 1 (1990).

\bibitem{kravchenko94-96} S.V.~Kravchenko, G.V.~Kravchenko,
  J.E.~Furneaux, V.M.~Pudalov, M.~D'Iorio, Phys. Rev. B {\bf 50}, 8039
  (1994);  S.V.~Kravchenko, W.E.~Mason, G.E.~Bowker, J.E.~Furneaux,
  V.M.~Pudalov, M.~D'Iorio, Phys. Rev. B {\bf 51}, 7038 (1995);
S.V.~Kravchenko, D.~Simonian, M.P.~Sarachik, W.~Mason, and
  J.E.~Furneaux, Phys. Rev. Lett. {\bf 77}, 4938 (1996).

\bibitem{abrahams01} E.~Abrahams, S.V.~Kravchenko, and M.P.~Sarachik,
  Rev. Mod. Phys. {\bf 73}, 251 (2001).

\bibitem{altshuler01} B.L.~Altshuler, D.L.~Maslov, and V.M.~Pudalov,
  Physics E {\bf 9}, 209 (2001).

\bibitem{kravchenko04}  S.V.~Kravchenko, and M.P.~Sarachik,
Rep. Prog. Phys. 67, 1 (2004).

\bibitem{pudalov04a} V.M.~Pudalov, cond-mat/0405315.

\bibitem{temp-dep-screen} F.~Stern, Phys. Rev. Lett. {\bf 44}, 1469
  (1980); A.~Gold and V.T.~Dolgopolov, Phys. Rev. B {\bf 33}, 1076
  (1986); S.~Das~Sarma, Phys. Rev. B {\bf 33}, 5401 (1986);
  S.~Das~Sarma and E.H.~Hwang, Phys. Rev. Lett. {\bf 83}, 164 (1999).

\bibitem{ZNA} G.~Zala, B.N.~Narozhny, and I.L.~Aleiner,
  Phys. Rev. B {\bf 64}, 214204 (2001); Phys. Rev. B {\bf 64}, 201201
  (2001); Phys. Rev. B {\bf 65}, 020201 (2002).

\bibitem{GM} I.V.~Gornyi and A.D.~Mirlin, Phys. Rev. Lett. {\bf 90}, 076801
  (2003); Phys. Rev. B {\bf 69}, 045313 (2004).

\bibitem{punnoose} A.~Punnoose and A.M.~Finkelstein,
  Phys. Rev. Lett. {\bf 88}, 016802 (2002);
  Science {\bf 310}, 289 (2005).

\bibitem{okamoto} T.~Okamoto, K.~Hosoya, S.~Kawaji, and A.~Yagi,
Phys. Rev. Lett. {\bf 82}, 3875 (1999).

\bibitem{kravchenko00} S.V.~Kravchenko, A.A.~Shashkin, D.A.~Bloore,
and  T.M.~Klapwijk, Solid State Commun. {\bf 116}, 495 (2000).

\bibitem{pudalov02} V.M.~Pudalov, M.E.~Gershenson, H.~Kojima, N.~Butch,
E.M.~Dizhur, G.~Brunthaler, A.~Prinz, and G.~Bauer,
Phys. Rev. Lett. {\bf 88}, 196404 (2002).

\bibitem{zhu} J.~Zhu, H.L.~Stormer, L.N.~Pfeiffer, K.W.~Baldwin,
and K.W.~West, Phys. Rev. Lett. {\bf 90}, 056805 (2003).

\bibitem{vitkalov01}   S.A.~Vitkalov, H.~Zgeng, K.M.~Mertes,
  M.P.~Sarachik, and T.M.~Klapwijk, Phys. Rev. Lett. {\bf
87}, 086401 (2001).

\bibitem{shashkin01} A.A.~Shashkin, S.V.~Kravchenko, V.T.~Dolgopolov,
  and T.M.~Klapwijk, Phys. Rev. Lett. {\bf 87}, 086801 (2001).

\bibitem{prus03} O.~Prus, Y.~Yaish, M.~Reznikov, U.~Sivan, and
  V.~Pudalov, Phys. Rev. B {\bf 67}, 205407 (2003).

\bibitem{pudalov04b} V.M.~Pudalov, M.E.~Gershenson, and H.~Kojima, in:
  {\it ``Fundamental Problems of Mesoscopic Physics. Interaction and
  Decoherence''}, edited by I.V.Lerner, B.L.Altshuler, and Y.Gefen
  (Kluwer, 2004), p.309.

\bibitem{shashkin05} A.A.~Shashkin, Physics-Uspekhi {\bf 48}, 129 (2005).

\bibitem{shashkin02} A.A.~Shashkin, S.V.~Kravchenko, V.T.~Dolgopolov,
  and T.M.~Klapwijk, Phys. Rev. B {\bf 66}, 073303 (2002).

\bibitem{proskuryakov02} Y.Y.~Proskuryakov, A.K.~Savchenko,
  S.S.~Safonov, M.~Pepper, M.Y.~Simmons, and D.A.~Ritchie,
  Phys. Rev. Lett. {\bf 89}, 076406 (2002);
 Y.Y.~Proskuryakov, A.K.~Savchenko,
  S.S.~Safonov, M.~Pepper, M.Y.~Simmons, D.A.~Ritchie, E.H.~Linfield,
  and Z.D.~Kvon, J. Phys. A {\bf 36}, 9249 (2003).

\bibitem{pudalov03}  V.M.~Pudalov, M.E.~Gershenson, H.~Kojima,
G.~Brunthaler, A.~Prinz, and G.~Bauer,
Phys. Rev. Lett. {\bf 91}, 126403 (2003).

\bibitem{anissimova05} S.~Anissimova, A.~Venkatesan, A.A.~Shashkin,
  M.R.~Sakr, S.V.~Kravchenko, and T.M.~Klapwijk, cond-mat/0503123.

\bibitem{shashkin03} A.A.~Shashkin, M.~Rahimi, S.~Anissimova,
  S.V.~Kravchenko, V.T.~Dolgopolov, and T.M.~Klapwijk,
  Phys. Rev. Lett. {\bf 91}, 046403 (2003); A.A.~Shashkin,
  S.V.~Kravchenko, V.T.~Dolgopolov, and T.M.~Klapwijk, J. Phys. A:
  Math. Gen. {\bf 36}, 9237 (2003).

\bibitem{MMR} G.M. Martin, D.L. Maslov, and M.Yu. Reizer,
Phys. Rev. B {\bf 68}, 241309(R) (2003).

\bibitem{lifshitz} I.M.~Lifshitz and A.M.~Kosevich, Zh.\ Eksp.\ Teor.\
Fiz.{\bf 29}, 730 (1955) [Sov. Phys. JETP {\bf 2}, 636 (1956)].

\bibitem{stamp} S.\ Curnoe and P.\ C.\ E.\ Stamp,
  Phys. Rev. Lett. {\bf 80}, 3312, (1998).

\bibitem{fowler}  M. Fowler and R. E. Prange, Physics {\bf 1}, 315 (1965);
 S. Engelsberg and G. Simpson, Phys. Rev. B {\bf 2}, 1657 (1970).

\bibitem{LuttiWard60} J. M. Luttinger and J. C. Ward
Phys. Rev. {\bf 118}, 1417 (1960).

\bibitem{AGD} A.A.~Abrikosov, L.P.~Gorkov, and I.E.~Dzyaloshinski,
{\em Methods of quantum field theory in statistical physics}, (Dover
Publications, New York,1963).

\bibitem{chubukov1}  A.V.~Chubukov and D.L.~Maslov, Phys. Rev. B {\bf 68}, 155113 (2003);
Phys. Rev. B {\bf 69}, 121102 (2004).

\bibitem{chubukov}
A.V.~Chubukov, D.L.~Maslov, S.~Gangadharaiah, and L.I.~Glazman,
Phys. Rev. B {\bf 71}, 205112 (2005); Phys. Rev. Lett. {\bf 95},
026402 (2005);
S.~Gangadharaiah, D.L.~Maslov, A.V.~Chubukov, and L.I.~Glazman,
Phys. Rev. Lett. {\bf 94}, 156407 (2005).

\bibitem{luttinger}  J.M. Luttinger, Phys. Rev. {\bf 121}, 1251 (1961).

\bibitem{ando} T.\ Ando, A.B.\ Fowler, and F.\ Stern,
 Rev.\ Mod.\ Phys.\ {\bf 54}, 437 (1982).
 
 \bibitem{MPW} In strong magnetic field and smooth disorder this is no longer true since
 the transport acquires adiabatic character duet classical memory effects,
 see e.g. A.D.~Mirlin, D.G.~Polyakov, and P.~W\"olfle, Phys. Rev. Lett. {\bf 80}, 2429 (1998) 

\bibitem{drag} I.V.\ Gornyi, A.D.\ Mirlin, and F. von Oppen,
Phys. Rev. B {\bf 70}, 245302 (2004).

\bibitem{rudin} A.\ M.\ Rudin, I.\ L.\ Aleiner, and L.\ I.\ Glazman,
\prb {\bf 55}, 9322 (1997).

\bibitem{Falko} G.~Zala, B.N.~Narozhny, I.L.~Aleiner, and
V.I.~Fal'ko,  Phys. Rev. B {\bf 69}, 075306 (2004).



\bibitem{MWA}
A.D.~Mirlin, E.~Altshuler, and P.~W\"olfle, Annalen der Physik {\bf 5},  281
  (1996).

\bibitem{CA} G.~Catelani and I.L.~Aleiner, JETP {\bf 100}, 331 (2005).

\bibitem{VaAl} M.G.~Vavilov and I.L.~Aleiner,
Phys. Rev. B {\bf 69}, 035303 (2004).

\bibitem{AAK} B.L.~Altshuler, A.G.~Aronov, and D.E.~Khmelnitsky,
  J. Phys. C {\bf 15}, 7367 (1982).

\bibitem{LM} T.~Ludwig and A.D.~Mirlin, Phys. Rev. B {\bf 69}, 193306 (2004).


\bibitem{DasSarma} V.M.~Galitski and S.~Das~Sarma,
Phys. Rev. B {\bf 70}, 035111 (2004).

\bibitem{aristov} This calculation is in fact very similar to the
analysis of the momentum dependence of the RKKY interaction between
the localized moments via the conduction electrons, see
D.N.~Aristov, Phys. Rev. B {\bf 55}, 8064 (1997).

\bibitem{footnote-i} Technically, this difference between the Coulomb interaction case 
and the short-range interaction case
can be traced back to the contribution
of $\Sigma_{00}$ represented by the first term in
the r.h.s. of Eq.~(\ref{product-singlet}).
For Coulomb interaction this term produces in the diffusive regime 
a $T\ln T$ contribution to the damping factor. 
In the ballistic regime, 
where the frequency dependence of screening becomes irrelevant
({\it cf.} the conductivity corrections, Refs.~\onlinecite{ZNA,GM})
and the Coulomb interaction reduces to a constant,
$V^\rho(i\omega_m,\bq)\simeq 1/2\nu,$ this term yields a  
subleading contribution as in the case
of the short-range interaction.
 

\bibitem{sav}
    E.A.~Galaktionov, A.K.~Savchenko, S.S.~Safonov, Y.Y.~Proskuryakov, L.~Li,
    M.~Pepper, M.Y.~Simmons, D.A.~Ritchie, E.H.~Linfield,
    and Z.D.~Kvon, in {\textit{Fundamental Problems of Mesoscopic Physics: Interactions and Decoherence}},
    edited by I.V.~Lerner, B.L.~Altshuler, and Y.~Gefen (Kluwer Academic Publishers, Dordrecht, 2004), p. 349.

\bibitem{footstoner}
For strong interaction $(F_0^\sigma)^2/(1+F_0^\sigma)\gg 1$
(in particular, in the vicinity of the ferromagnetic instability $F_0^\sigma\to -1$)
an additional regime, $(1+F_0^\sigma)/(F_0^\sigma)^2\ll 2\pi T\tau \ll 1,$ appears in the diffusive range of $T$.

\end{thebibliography}
\end{document}